\documentclass[journal=jacsat,manuscript=article]{achemso}
\usepackage{longtable}
\usepackage{booktabs}
\usepackage{multirow}
\usepackage{amsmath}
\usepackage{subfigure}
\usepackage[usenames,dvipsnames]{color}
\usepackage{soul}
\usepackage{bm}
\usepackage{rotating}
\usepackage[version=3]{mhchem} 
\usepackage{chemfig}

\usepackage{amssymb}
\usepackage{titlecaps}
\usepackage{xcolor}
\usepackage{tikz}
\usepackage{hyperref}
\usepackage{xr}
\usepackage{import}
\usepackage[normalem]{ulem}

\setchemfig{atom sep=2em}
\SectionNumbersOn

\author{Valerii Andreichev} \altaffiliation{These authors
  contributed equally} \affiliation{Department of Chemistry, University of Basel,
  Klingelbergstrasse 80, CH-4056 Basel, Switzerland}

\author{Sena Aydin} \altaffiliation{These authors
  contributed equally} \affiliation{Department of Chemistry, University of Basel,
  Klingelbergstrasse 80, CH-4056 Basel, Switzerland}
  
\author{Kai T\"opfer} \alsoaffiliation{Department of Chemistry,
  University of Basel, Klingelbergstrasse 80, CH-4056 Basel,
  Switzerland} \affiliation{Department of Biology, Chemistry and
  Pharmacy, Freie Universität Berlin, Arnimallee 22, Berlin, DE-14195
  Germany}

\author{Markus Meuwly} \affiliation{Department of Chemistry,
  University of Basel, Klingelbergstrasse 80, CH-4056 Basel,
  Switzerland} \email{m.meuwly@unibas.ch} \author{Luis Itza
  Vazquez-Salazar}\alsoaffiliation{Department of Chemistry, University
  of Basel, Klingelbergstrasse 80, CH-4056 Basel, Switzerland}
\affiliation{Institute for Theoretical Physics, Heidelberg University,
  Philosophenweg 12, DE-69120 Heidelberg, Germany}

\email{liitzavazquezs@gmail.com}

\title{Design, Assessment, and Application of Machine Learning Potential Energy Surfaces}

\begin{document}

\begin{abstract}
Potential Energy Surfaces (PESs) are an indispensable tool to
investigate, characterise and understand chemical and biological
systems in the gas and condensed phases. Advances in Machine Learning
(ML) methodologies have led to the development of Machine Learned
Potential Energy Surfaces (ML-PES) which are now widely used to
simulate such systems. The present work provides an overview of
concepts, methodologies and recommendations for constructing and using
ML-PESs. The choice of topics is focused on practical and recurrent
issues to conceive and use such model. Application of the principles
discussed are illustrated through two different systems of
biomolecular importance: the non-reactive dynamics of the
Alanine-Lysine-Alanine tripeptide in gas and solution phases, and
double proton transfer reactions in DNA base pairs.
\end{abstract}

\section{Introduction}
The use of Machine learning (ML) methods is changing the way we do
science. Specifically, these new technologies become part of the
standard toolbox for studying physical and chemical processes. This
has been recently recognised by awarding the 2024 Nobel Prizes in
Physics and Chemistry to G. Hinton/J. Hopfield and
D. Hassabis/J. Jumper/D. Baker. This is not surprising, considering
that the use of ML methodologies helps scientists generate new
hypotheses, analyze large quantities of data, speed up simulation
times, and gain insights from data that cannot be obtained using
traditional scientific methods.\cite{wang2023scientific} In addition,
the increase in computational power (Moore's
Law)\cite{moore1998cramming} and continuous investment in artificial
intelligence (AI)\cite{quid2025_ai_investment} stimulates the adoption
and application of ML methodologies in the natural sciences. In
Chemistry and Physics, ML starts to make a clear impact as evidenced,
e.g., through the discovery of a new phase transition in liquid
hydrogen\cite{cheng2020evidence} by using ML-aided simulations or the
repositioning of Halicin, a drug originally designed to treat
diabetes, as an antibiotic.\cite{zhavoronkov2019deep}\\

\noindent
The impact of the use of ML methodologies has also been reflected in
the simulation of the dynamics and reactivity of molecules in the gas
phase and in solution. In particular, atomistic simulations using
machine-learned potential energy surfaces (ML-PESs) have been
significantly impacted over the past 10 years. For example, the winner
of the 2020 ACM Gordon Bell prize simulated a water box with
approximately 4M molecules ($\approx 12$M atoms) and a solid-state
system of Cu with 100M atoms\cite{lu202186} using the DeepMD
model\cite{e2018dpmd}. Similarly, ML-PESs have been used in atomistic
simulations for biomolecular
importance\cite{unke2024biomolecular,zaverkin2025performance}. The
currently largest such simulation was that of the HIV capsid with 44M
atoms using the Allegro model\cite{musaelian2023learning} have been
reported.\cite{kozinsky2023scaling} However, such impressive system
sizes are still small compared to what can be handled using empirical
energy functions (billions of atoms for several million
steps)\cite{kadau2006molecular,germann2008trillion,shibuta2017heterogeneity,casalino2022breathing,santos2024breaking,ugarte2025scaling}. Nevertheless,
the accuracy obtained using an ML-PES is remarkable as it is closer to
\textit{ab initio} MD simulations, which are usually unfeasible for
systems larger than a few hundred atoms, depending on the level of
theory used. \\

\noindent
Within the broader physical chemistry community, ML-PESs together with
established molecular dynamics (MD) suites and methods becomes a
standard
procedure\cite{lahey2020,gastegger2021mlsolv,inizan2023scalable,MM.pycharmm:2023,zinovjev2023electrostatic,zinovjev2024emle,kalayan2024neural}
Likewise, performance and stability challenges for ML-PESs have been
established that explicitly target biomolecular
systems.\cite{poltavsky2025crash} Such studies are needed to pave the
way for treating larger systems using mixed methods (i.e. Machine
Learning/Molecular Mechanics (ML/MM)). Conversely, progress in other
areas of chemistry, like the appearance of foundational models, which
are any model trained on broad data that can be adapted to a wide
range of tasks \cite{bommasani2021opportunities}, in materials
chemistry\cite{batatia2023foundation,pyzer2025foundation},
drug-discovery\cite{ple2025foundation} and chemistry in
general\cite{yuan2025foundation,choi2025perspective} is evidence that
there are still many areas of opportunity that have not been explored
for ML-PESs. \\

\noindent
To set the stage for the present work, it is useful to clarify and
define some terminology. In the literature, machine learning-based
potential energy surface (ML-PES), force field (ML-FF), machine
learning potential (MLP) or machine learning interatomic potential
(ML-IP), among others, can be found and used interchangeably. In the
following, MLP is disfavoured because it can be confused with
"multilayer perceptron", which is extensively used in the
ML-literature at large. An "empirical energy function" (EEF), on the
other hand, is a parametrized energy expression for which parameters
are determined through minimization procedures, typically using
least-squares fitting procedures\cite{cgenff:2012} or more advanced
techniques, such as graph
NNs.\cite{wang2022end,takaba2024machine,wang2024espalomacharge} In the
following, ML-PES will be used throughout as a generic term for a
neural network or kernel-based representation of reference data from
quantum chemical calculations.\\

\noindent
Historically, it is of interest to note that J. E. (Lennard-)Jones
wrote\cite{jones:1924} already in 1924 (italics added): "Until our
knowledge of the disposition and motion of the electrons in atoms and
molecules is more complete, we cannot hope to make a direct
calculation of the nature of the {\it forces} called into play during
an encounter between molecules in a gas. It is true that [..] Debye
[..]investigated the nature of the field in the neighbourhood of a
hydrogen atom [..] and has shown how the pulsating field gives rise on
the whole to a force of repulsion, as well as one of attraction on a
unit negative charge. But it is difficult to see how this work can be
extended to more complex systems.[..] One such method is to {\it
  assume a definite law of force}, and then by the methods of the
kinetic theory to deduce the appropriate law of dependence of the
viscosity of a gas on temperature." Notably, Lifson and Warshel
adopted similar language based on their focus on vibrational
spectroscopy and the related "force constants" in their 1968 landmark
publication.\cite{warshel:1968} They write "[..]A set of functions
with such optimized parameters will be referred to as a consistent
force field [..]" Obviously, both, Lennard-Jones and Lifson/Warshel
thought of "force" instead of "energy", although the expressions that
were parametrized described how the total energy changes with
geometry.\\

\noindent
Despite the growing excitement for the standard application of
ML-PESs, it is still technology in development and as a community we
need to learn how to best and reliably employ them effectively for the
simulation of molecular systems. Important aspects concern the
validity and reproducibility of simulation using ML-PESs. A reason for
this is the fact that the process of creating and using an ML-PES
requires several steps\cite{MM.rev:2023} In addition, for each of
those steps, there are several options regarding software,
methodologies and possible problems. In most cases, there is not a
single-best solution that adapts to the different needs of
practitioners.\\

\noindent
Keeping this in mind, it appears to be timely to discuss "best
practices" for generating and using ML-PES for valid
simulations. Early efforts in this regard have been made for chemistry
in general\cite{artrith2021best}, for the selection of ML
potentials\cite{pinheiro2021choosing}, and for the validation of
ML-PES\cite{morrow2023validate}. In addition, common guidelines from
the molecular dynamics community regarding the sharing of
data\cite{abraham2019sharing} or the construction of reproducible
simulations\cite{elofsson2019ten}, as well as contributions from the
material science
community\cite{scheffler2022fair,ghiringhelli2023shared}, provide
valuable guidelines. \\

\noindent
In the following, important aspects for generating ML-PESs are
considered and recommendations are formulated for successful model
building. Starting with the construction of an ML-PES, followed by the
validation of a trained model, improvements to an ML-PES, and the
practical aspects of using an ML-PES. Two concrete examples of
biomolecular interest are then presented to illustrate the
recommendations. The first corresponds to a non-reactive system: the
ALA-LYS-ALA tripeptide, and the second concerns the proton transfer
process in DNA base-pairs. Finally, open questions, challenges, and
perspectives are formulated.\\

\section{Construction of ML-PESs}
As mentioned in the introduction, constructing an ML-PES is a
multi-step procedure.\cite{MM.rev:2023} Starting from the definition
of the problem of interest or the phenomenon to be described, the next
natural step is to select a model to study the problem of
interest. The underlying model - typically a neural network (NN) or
kernel-based technique needs to be decided on. There is by now a
multitude of different options available. A non-exhaustive list
includes, but is not limited to,
SchNet\cite{schuett2017schnet,schutt2018schnet},
PhysNet\cite{MM.physnet:2019},PaiNN\cite{schuett2021painn},
SpookyNet\cite{unke2021spookynet}, Nequip\cite{batzner2022nequip},
MACE\cite{batatia2022mace},
So3krates\cite{frank2022so3krates,frank2024sokrates},
allegro\cite{musaelian2023learning}, ANIKEN-ME
models\cite{smith2017ani},
FCHL\cite{faber2017prediction,lilienfeld2020fchl},
SGLD\cite{chmiela2023accurate}, reproducing
kernels\cite{rabitz:1996,hollebeek.annrevphychem.1999.rkhs,unke2017toolkit},
and KerNN\cite{mm.kernn:2025}, among many others. The following
discussion is agnostic to any of these particular implementations and
models. Rather, for the most part, the focus is on conceiving,
testing, improving, validating and using ML-PESs for practical
applications.\\

\begin{figure}
    \centering
    \includegraphics[width=\linewidth]{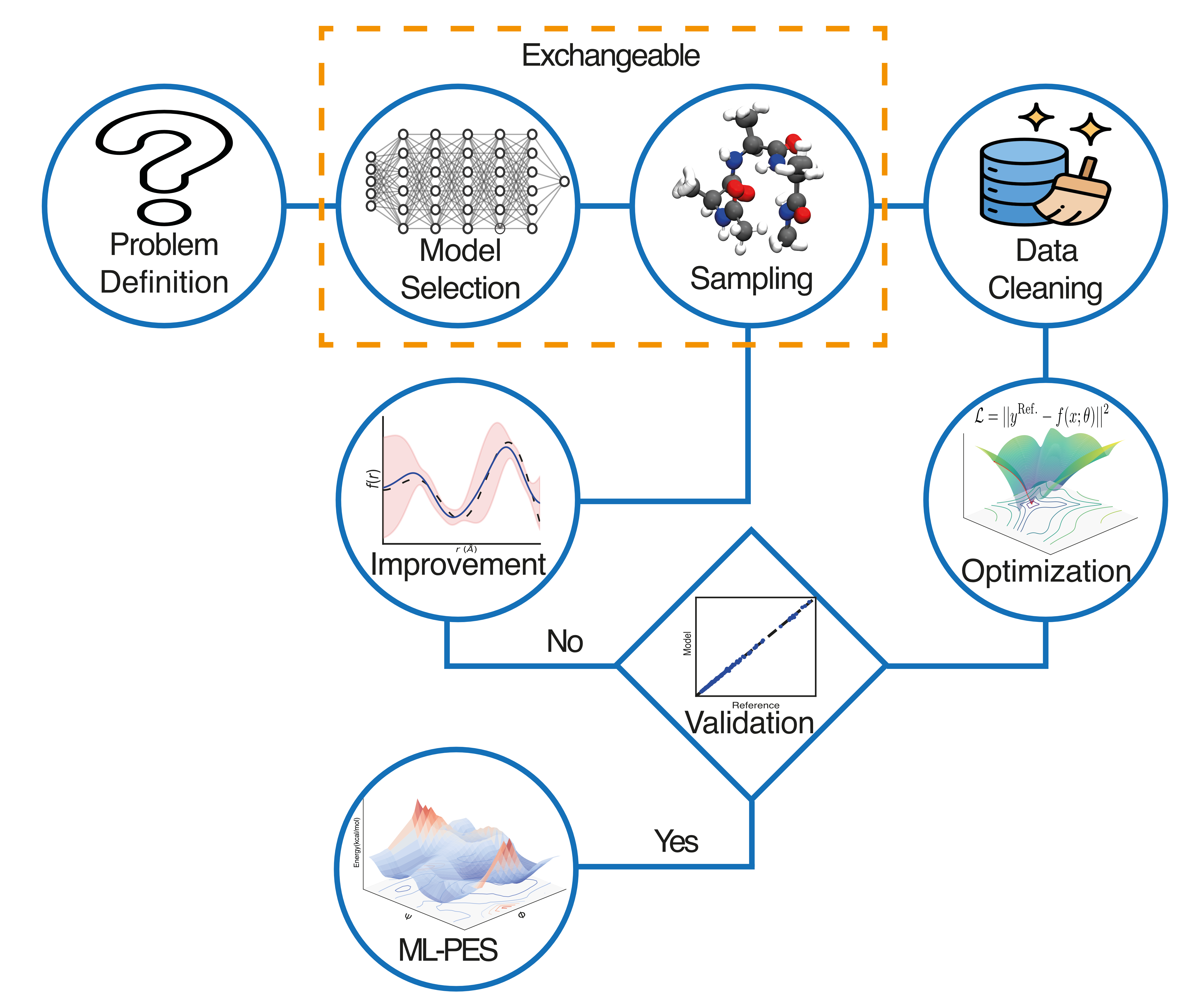}
    \caption{\textbf{Construction of a ML-PES} The process begins by
      defining the problem and selecting an appropriate ML model. This
      is followed by an iterative cycle of data generation (sampling),
      cleaning, training, validation, and refinement. Once validated,
      the model can be employed for molecular simulations.}
    \label{fig:constr_ML-PES}
\end{figure}

\noindent
Once the problem to solve has been defined and a model was selected,
the next step is to obtain reference data to train on. This can be
done by either using previously generated datasets (if available) or
starting from scratch using appropriate sampling strategies. Next, the
model needs to be trained by minimising a loss function. This step
finishes once the model is considered to be converged or it has
reached the desired accuracy. The model then needs to be verified and
validated. For each of the previously described steps, numerical tools
and procedures are available. As a last step, the robustness of a
trained model needs to be determined. A graphical summary of the steps
to generate an ML-PES is shown in Figure \ref{fig:constr_ML-PES}. In
the following, each step will be described separately and
recommendations are provided. \\

\subsection{Models for Machine Learned Potential Energy Surfaces}
One of the first considerations for building an ML-PES concerns the
machine learning model to be used. For representing inter- and
intramolecular interactions two main avenues have emerged over the
past 10 years: kernel- and neural network-based approaches. In the
following, the focus is on NN-representations. However, for certain
applications - for example, for small molecules - kernel-based methods
have been applied successfully. A non-exhaustive list of kernel models
includes Symmetric Gradient Domain Machine Learning
(sGDML)\cite{chmiela2023accurate} or Reproducing kernel Hilbert space
(RKHS) \cite{ho96:2584,MM.rkhs:2017} with applications to high-energy
reactions.\cite{MM.heh2:2019,MM.heh2:2023,MM.heh:2025} The interested
reader is referred to previous reviews on the
topic.\cite{rupp2015machine,pronobis2020kernel,deringer2021gaussian,pinheiro2023kernel}\\

\noindent
Considerations for choosing a specific ML-model should (also) be based
on the nature of the problem to be solved.\cite{medina2021rationality}
Examples for ''problems'' include i) chemical reactivity, ii) dynamics
in the gas or condensed phase, iii) spectroscopic investigations, iv)
dynamics on surfaces, or v) ligand binding studies, to name a
few. Before selecting a particular NN-architecture, it is useful to
consult prior applications and known limitations of specific types of
implementations. In addition, for the NN-architecture chosen, the
likely amount of training data required should be estimated. Despite
their greater flexibility and reduced inductive bias, computationally
intensive architectures may not always offer advantages over simple
models such as fully connected neural networks. The selection of a NN
for constructing and using a ML-PES should be based mainly on two
aspects: the descriptor of the molecular system and the NN
type/architecture.\\

\subsection{Features, Descriptors and Representations}
A descriptor is a numerical representation that allows the algorithm
to detect patterns.\cite{keith2021combining} The purpose of a suitable
representation is to relate input data (e.g. structure) with the
property to be predicted (e.g. energy).\cite{musil2021physics} The
task to represent a molecule for its use within an ML model has
multiple possible solutions. Therefore, this is an area of active
research\cite{wigh2022review,raghunathan2022molecular,jones2023molecular}.\\

\noindent
An ideal descriptor for training and representing a ML-PES has the
following
characteristics\cite{knoll2015descr,huo2022unified,uhrin2021descr,rupp2022descr,MM.rev:2023,jones2023molecular}:
it is 1) \textit{Invariant} to rotational, reflection, translational,
and permutational symmetry. 2) \textit{Unique} to warrant a one-to-one
mapping between the input structure and the output. 3) {\it Complete}
to ensure the injectivity of the representation. 4)
\textit{Descriptive} to provide sufficient information to describe the
molecular system accurately. 5) \textit{Continuous and Differentiable}
which is fundamental for obtaining derived properties such as forces
($F_{i} = \nabla_{i} E_{i}$) or Hessians ($H_{i} =
\nabla_{i}^{2}E_{i}$) and 6) \textit{Simple} being computationally
efficient and generalizable across multiple chemical systems. It
should, however, be noted that to the best of the authors' knowledge
no such "ideal descriptor" has been developed and used so far for
ML-PESs.  \\

\begin{figure}
    \centering
    \includegraphics[width=0.9\linewidth]{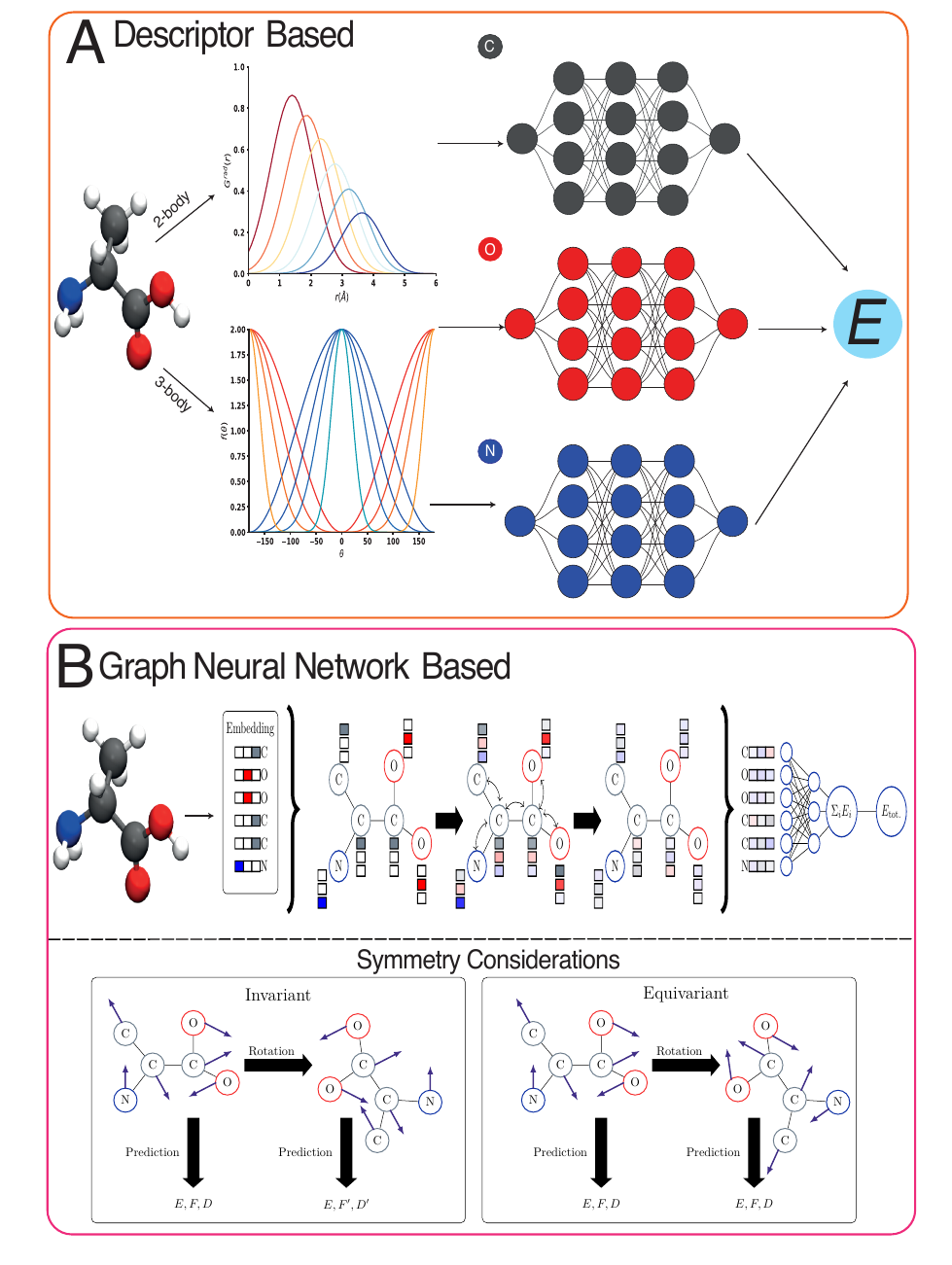}
    \caption{\textbf{Different types of neural networks} Panel A:
      Models with fixed descriptors represent molecules using
      predefined two- and three-body functions. Separate multilayer
      perceptrons predict atomic energies, which are summed to obtain
      the total molecular energy. Panel B: Graph neural network (GNN)
      models represent molecules as graphs of atoms (nodes) and bonds
      (edges). Each atom in the molecule is described with an initial
      embedding vector. Atomic embeddings are iteratively updated
      through message passing and passed to a readout function to
      predict atomic energies. The lower panels illustrate symmetry
      conservation: invariant models preserve scalar properties (e.g.,
      energy) under rotation, while equivariant models additionally
      conserve vectorial quantities (e.g., forces, dipoles).}
    \label{fig:NN_models}
\end{figure}

\noindent
Descriptors can be classified into two categories: predefined and
learnable. {\it Predefined descriptors} use fixed functional forms to
describe interatomic interactions and were among the first to be
explored following the pioneering work of Behler in
2007\cite{behler2007generalized}. The most recognized predefined
descriptors are the atom-centered symmetry functions
(ACSF)\cite{behler2011acsf,behler2015constructing} in which an atom
type in a molecular system is described with radial (2-body) and
angular (3-body) terms as shown in Figure \ref{fig:NN_models}A. The
resulting chemical (atomic) environment is then passed to a fully
connected NN that returns the contribution of each atom to the total
energy. This strategy presents problems for multicomponent systems due
to the unfavourable increase in the number of ACSF with the number of
elements. Alternatives to ACSF include, but are not limited to,
ANAKIN-ME (Accurate NeurAl networK engINe for Molecular Energies;
ANI), \cite{smith2017ani}, tensormol\cite{yao2018tensormol}, deep
potential\cite{e2018dpmd,wang2018deepmd} and smooth-deep potential
representation\cite{zhang2018end}, descriptors based on neighbourhood
density functions\cite{khorshidi2016amp,unke2018reactive}, moment
tensor potential\cite{shapeev2016moment}, atomic cluster expansion
(ACE)\cite{drautz2019atomic} and
variations\cite{cheng2024cartesian}.\\

\noindent
A challenge in the use of predefined descriptors lies in selecting the
hyperparameters of the underlying functional forms. This choice is
highly system-dependent and, usually, requires expert knowledge. Poor
choices can result in an incomplete description of the system and lead
to degenerate energy predictions with respect to molecular
geometry.\cite{ceriotti2020incomplete,goedecker2022fingerprints}
Consequently, hyperparameter values must be carefully benchmarked
before a model can be reliably used in simulations. General
guidelines, however, do exist for common quantities. For example, when
choosing the cutoff radius for radial functions, values that are too
small should be avoided, as they introduce artifacts and inaccuracies
in force calculations near the cutoff.\cite{behler2011acsf} The
specific cutoff function should also be selected with computational
considerations in mind (e.g., speed, memory, accuracy). Other
parameters, such as the number of basis functions or the order of
$n-$body interactions, should be determined based on convergence of
the training of the model and the desired level of accuracy for
it. Detailed strategies for tuning hyperparameters in different
predefined descriptors can be found in Ref. \citenum{behler2021four}
and \citenum{tokita2023train} for ACSF,
Ref. \citenum{bochkarev2022efficient} for ACE, and
Ref. \citenum{pan2024training} for the Deep Potential framework, with
additional examples available for reactive systems. \\

\noindent
{\it Learnable descriptors} (or end-to-end descriptors) do not rely on
predefined functions to characterise a molecular system. Instead, a
suitable representation is directly trained from fitting to
data.\cite{unke2021machine} Such descriptors are rooted in graph
neural networks (GNNs)\cite{scarselli2009gnn}, particularly
message-passing neural networks (MPNNs)\cite{gilmer2017mpnn}, whereby
a molecule is represented as a graph with atoms as nodes and bonds as
edges (Figure \ref{fig:NN_models}B). Each node is associated with an
embedding that may encode atomic properties (e.g.,
charges\cite{coley2017convolutional}) or be randomly initialized,
while edges typically represent bond information such as bond order or
interatomic distances. Within this framework, the MPNN operates in two
stages (see Figure\ref{fig:NN_models}B): (i) a message-passing step,
in which node embeddings are iteratively updated through messages
generated by a vertex update function that aggregates information from
neighboring nodes, and (ii) a readout step, which applies a function
to the updated embeddings to return the final atomic energy
contributions. \\

\noindent
Recent approaches extend the representation of molecules from purely
topological graphs to geometric graphs, where nodes are described not
only by embedding vectors but also by their spatial
coordinates.\cite{velivckovic2023everything} These geometric graphs
form the basis of geometric deep
learning\cite{bronstein2017geometric,bronstein2021geometric}, a
rapidly growing branch of machine learning with numerous applications
in chemistry and drug discovery.\cite{atz2021geometric} Geometric
graphs are particularly well suited for representing molecular systems
as their physical invariances can be explicitly incorporated. Building
on this idea, new descriptors that are equivariant (covariant) to the
geometric transformations (translation, rotation, and reflection) of
the Euclidean group $E(3)$\cite{velivckovic2023everything} have
emerged. These descriptors have gained considerable popularity thanks
to the routine calculation of vectorial and tensorial quantities for
MD simulations (Figure \ref{fig:NN_models}B). Incorporating geometric
information into GNN-based descriptors enhances predictive accuracy,
generalization, and data
efficiency\cite{MM.rev:2023,duval2023hitchhiker}, but at the expense
of increased computational cost and reduced
interpretability. Additionally, it should be noted that some learnable
descriptors depend on the NN-architecture, and special operations are
required for their use.  \\

\noindent
Learnable descriptors are the product of an optimisation process,
which, however, still requires user-defined hyperparameters. One of
them concerns the basis functions used to encode the spatial
coordinates of the system, which can be classified into invariant or
equivariant types depending on which molecular symmetries
(translational, rotational, and permutational) are encoded.  Invariant
approaches were the first developed using radial basis functions
(RBFs), analogous to ACSF for two body terms.  Models like
SchNet\cite{schutt2018schnet} and PhysNet\cite{MM.physnet:2019} use
RBFs depending on internuclear separations. Improvements to invariant
approaches introduced three-body angular terms using combinations of
RBFs with spherical Fourier–Bessel
functions\cite{gasteiger2019directional} and, later, four-body
dihedral terms\cite{gasteiger2021gemnet} by using evolving Bessel
polynomials. Equivariant basis functions, in contrast, employ
spherical harmonics to explicitly enforce covariance with respect to
the Euclidean group $E(3)$. Notice that an equivariant descriptor can
also be obtained without the use of an equivariant basis function by
modifying the message passing step\cite{schuett2021painn} to
incorporate directional information.\\

\noindent
A second hyperparameter to be chosen is length of the embedding
vector. This vector needs to be large enough to distinguish atomic
environments\cite{duval2023hitchhiker} but not so large to risk
overfitting\cite{barros2018hipnn,MM.rev:2023}; multiples of 32 are
typically used for hardware efficiency, with values between 64 and 128
being most common. Finally, the cutoff radius defines the interaction
range considered during descriptor construction and message passing.
Larger cutoff values capture more neighbors and information but at a
higher computational cost, whereas smaller cutoffs risk omitting
relevant interactions. Notably, equivariant descriptors can remain
accurate even with reduced cutoff values\cite{schuett2021painn}, as
geometric biases provide additional structural information.\\

\noindent
\textbf{Recommendations:} The choice between pre-defined or learnable
descriptors requires weighing multiple factors, including the system's
composition, the desired accuracy, and inference time. Predefined
descriptors offer advantages such as generally shorter inference times
and simplicity in their definition. However, they require expert
knowledge and extensive benchmarking before using them for novel
systems. Conversely, learnable descriptors are more flexible, provide
higher accuracies, and allow the natural inclusion of symmetry
constraints.  In particular, equivariant descriptors offer the
advantage of improved prediction of vectorial and tensorial quantities
and require low amounts of data in comparison with their invariant
competitors. It is noted that equivariance can also be achieved by
using data augmentation strategies during the
training\cite{langer2024probing} although such an approach typically
increases training times but can shorten inference times due to the
simpler NN-architecture\cite{brehmer2024does}.  Regardless of its type, learnable
descriptors are associated with higher computational costs, increased
memory usage, reduced interpretability, and longer inference
times. Comparative benchmarks between predefined and learnable
descriptors are available for materials
science\cite{zuo2020performance,leimeroth2025machine} and for
biomolecular simulations\cite{poltavsky2025crash}.  \\

\noindent
In addition to the previous requirements, other considerations may
also be taken into account.  As an example, for small, reactive
systems at low temperatures, the long-range part of the PES is crucial
for meaningful dynamics
studies.\cite{MM.heh2:2019,MM.heh2:2023,MM.heh:2025} Hence, the type
of model -- NN- vs. kernel-based -- needs to handle long-range
interactions in a sufficiently stable and accurate manner. Reproducing
kernel Hilbert space (RKHS) techniques\cite{ho96:2584,MM.rkhs:2017}
have emerged as one suitable approach for this. While standard NN
representations are known to either fail\cite{unke2021machine},
require large amounts of reference data\cite{MM.criegee:2023} or need
to explicit account for
them\cite{zubatyuk2019accurate,ko2021fourth,anstine2023machine,shaidu2024incorporating}. Conversely,
mixed representations, for example, 1d-RKHSs as features together with
a small NN \cite{mm.kernn:2025,MM.heh:2025} or hybrid
descriptors\cite{dezaphie2025designing} are interesting
alternatives. \\

\subsection{Neural Network Architecture}
The second key aspect in generating and using ML-PESs is the selection
and design of the neural network (NN) architecture. In the following,
"architecture" comprises parameters such as the number of layers, the
number of nodes, the type of activation function, or the type of NN
(such as discriminative NN or variational autoencoder). While certain
architectural details are tied to the choice of descriptor, see above,
other considerations require discussion. The first is the selection of
one of the key components of any NN-architecture: the activation
function which introduces ``nonlinearity'' to the model, enabling it
to learn relationships between input and output. General requirements
for activation functions are discussed in
Ref. \citenum{dubey2022activation}, with a comprehensive survey
provided in Ref. \citenum{kunc2024three}. For ML-PESs the selection of
an activation function should be made considering that the output
space of the NN must be continuous and smooth to allow the calculation
of derivatives for the calculation of forces (first derivative) or
Hessians (second derivative). Hence, smooth functions such as the
$Swish=\frac{x}{1+e^{-\beta x}}$ are
preferred\cite{duval2023hitchhiker} over alternatives with hard
cutoffs (e.g.  ReLu) as those can introduce discontinuities in the
derivatives.\cite{behler2021four} \\

\noindent
A further consideration when selecting a NN-architecture is the number
of layers. In general, a NN can be seen as a collection of input,
hidden, and output layers. The label of 'deep' NN is assigned when the
NN consists of three or more hidden
layers.\cite{prince2023understanding} Ideally, the number of layers
should not be too large: the universal approximation
theorem\cite{gybenko1989approximation,hornik1989multilayer,hornik1991approximation}
formally proves that a NN with two layers and a sufficiently large
feature space can approximate any continuous function on a compact
subset of the real line.\cite{prince2023understanding} However, in
practice, deeper models are used because they offer better performance
and parameter efficiency, though at the cost of greater susceptibility
to overfitting.\cite{behler2017first} As a general recommendation, it
is suggested to use 2 to 5 hidden layers for models aimed at working
with learnable descriptors. This recommendation can be extended to
fully connected NNs that are components of GNNs.\\

\noindent
The notion of "depth" is less relevant for GNN-architectures as those
usually involve several blocks consisting of multiple NNs that include
message passing (see above). Therefore, for GNNs, it is convenient to
consider the number of message-passing steps also referred to as
interaction layers. In an MPNN on each interaction layer, node
embeddings are updated by aggregating messages from neighboring nodes,
a process analogous to applying a convolutional operation over the
molecular graph.\cite{reiser2022graph} The number of interaction
layers needs to be carefully selected: Too few layers may not be able
to convey the message efficiently, and too many layers cause
over-smoothing. This can render node embeddings indistinguishable and
lead to over-squashing, whereby information is excessively compressed
through bottleneck edges.\cite{giraldo2023trade} Empirical evidence
suggests that 3–5 message-passing steps strike the best balance for
ML-PESs.\cite{MM.rev:2023}\\

\noindent
As concrete examples, the ANI\cite{smith2017ani} and
MACE\cite{batatia2022mace} architectures are briefly compared. The ANI
architecture achieves computational efficiency by using fixed,
hand-crafted local descriptors and small feed-forward neural networks
to predict individual atomic energies, avoiding the costly
message-passing and tensor operations used in models like MACE. This
design makes ANI fast and scalable, with linear computational cost,
but it limits its ability to capture many-body and directional
interactions. In contrast, MACE employs an $E(3)-$equivariant
message-passing framework that dynamically learns atomic
representations and encodes higher-order geometric correlations,
resulting in much greater physical fidelity and accuracy. However,
these richer interactions and tensor operations make MACE
substantially more computationally demanding than ANI.\\
  
\noindent
Other important aspects to consider when choosing a neural network
architecture include floating-point precision and the programming
language. Most modern ML algorithms operate with single-precision
(32-bit) arithmetic; however, in MD simulations this can be
problematic. Recent studies\cite{MM.acc:2024} have shown that models
trained in single precision produce unreliable derivatives, yielding
rougher ML-PES surfaces compared to those obtained with double
precision (64-bit). The programming language used to implement the
model directly affects execution speed\cite{heer2023speed}, energy
consumption\cite{pereira2017energy,marini2024green}, portability, GPU
compatibility, community support, and integration with legacy MD
codes. Broadly speaking, languages can be divided into compiled (e.g.,
Fortran, C, C++) and interpreted (e.g., Python, Matlab) programming
languages, with intermediate approaches such as Julia.\\

\noindent
Compiled languages generally deliver higher performance, lower energy
consumption, and direct access to GPU programming frameworks (e.g.,
CUDA), while also benefiting from long-term integration with
established MD suites. This comes at the expense of a more complex
syntax, the requirement to explicitly handle hardware resources, and
hardware-dependent compilation, which reduces portability. Compiled
languages have been used primarily for constructing models based on
ACSF, with notable examples including RuNNer\cite{runner2025},
RUBNNet4MD\cite{rubnn2025}, and n2p2\cite{n2p2}.\\

\noindent
Interpreted languages, by contrast, are slower and less energy
efficient, but their ease of use, portability, and large user
communities have made them the default in machine learning—and in
particular, Python has emerged as the dominant choice for ML-PES
development. Python’s rich ecosystem of libraries lowers the entry
barrier by simplifying model construction and providing GPU
acceleration through wrappers to compiled frameworks. The most widely
used packages are TensorFlow\cite{tensorflow2015-whitepaper},
PyTorch\cite{paszke2019pytorch}, and JAX\cite{jax2018github}, and the
choice among them typically depends on user preference and
convenience. Nevertheless, developers must pay careful attention to
versioning, as frequent updates in these libraries may introduce
changes that render earlier implementations obsolete. \\
 
\noindent
\textbf{Recommendations:} The choice of which NN-architecture to use
for representing a PES depends on the user needs for inference
performance (speed) and its accuracy in reproducing the reference
data. Generally speaking, comparably less accurate models such as
ANI\cite{smith2017ani} are much faster than highly accurate models
such as MACE.\cite{batatia2022mace} Naturally, the choice also depends
on the application which, in turn, determines the expected number of
ML-PES model evaluations (e.g. MD simulations). For representing high
level reference data (e.g. CASPT2 or CCSD(T)) the underlying
NN-architecture needs to be trainable to sufficient accuracy to not
offset the gain in accuracy of the electronic structure method
relative to a lower level method (e.g. DFT). Other special
considerations to make when, e.g., using equivariant NNs where the
atomic forces are predicted by a NN itself and not \textit{via} energy
derivatives through backpropagation. Force predictions \textit{via} a
ML-PES might be faster than computing the energy derivative, but
energy conservation is lost when used for $NVE$ simulations.\\

\subsection{Data Generation and Cleaning}
Depending on the application(s) in mind, data sets used for training
NN-based ML-PESs need to cover a wide range of geometries. Similarly,
for quantitative dynamics studies, the underlying quantum chemical
method needs to fulfill the highest standards, including approaches
such as full configuration interaction (FCI) with a large basis
set. However, this can only be afforded for systems containing a small
number of electrons. If, due to the chemical system, a lower level of
electronic structure method needs to be employed, one recommendation
is to carry out test calculations at a higher level of theory for
error estimation. As an example, if only calculations with triple zeta
basis set can routinely be carried out, a few calculations with a
larger basis set should be carried out as verification. \\

\noindent
For applications to medium/large(r) molecules, such as those of
biological interest, only medium-level \textit{ab-initio} electronic
structure methods (second order M{\o}ller-Plesset (MP2)) or density
functional theory (DFT) approaches can be used. Here it is recommended
to perform a few calculations at a higher level of theory (e.g. local
coupled cluster) for selected structures to obtain error estimates of
the quantum chemical method itself.\\

\noindent
The cornerstone for constructing a ML-PES is the data used to train
it. For an ML method, being trained with the necessary information is
as essential as having the appropriate reactants for a chemical
synthesis.\cite{fourches2010trust} In addition, the practitioner
should consider that ML models do not assume a functional
form. Therefore, the model would derive it from the reference
data.\cite{behler2021four} An open challenge in the field concerns the
{\it number} of data points and the {\it geometrical structures
  chosen} to cover the configurational space of a system of interest
to obtain a robust ML potential. This problem arises from the fact
that the description of a PES heavily depends on the system's size and
the molecular systems' structural complexity. Complementary, obtaining
complete coverage of chemical and conformational space for the
construction of ML-PES is an impossible task; therefore, biases in
databases are inevitable.\cite{wang2025design}\\

\noindent
Recent progress in tackling those challenges has been made for small
molecules\cite{MM.tl:2022,MM.tl:2025} on the problem of quantum
tunnelling splittings, which requires the construction of a PES at a
high level of theory. Obtaining high-quality data for molecules of the
size of tropolone (9 heavy atoms) at the CCSD(T)/aug-cc-pVTZ level of
theory is very challenging or even
impossible\cite{nandi2023ring}. Therefore, a solution based on using
transfer learning to improve a base ML-PES constructed at a modest
level of theory (MP2) to the required level of theory (CCSD(T)) using
a minimal number of high-level calculations selected using an
algorithm based on farthest-point-sampling\cite{garcke:2023} was
employed. The results indicate that for the accurate reproduction of
measured tunnelling splitting, a small number of samples (25) taken
from the instanton path was enough. Nevertheless, in cases where the
process to be described is not well-characterised, generating a
training dataset for an ML-PES is usually an iterative process. \\

\noindent
An alternative strategy for constructing ML-PES is to perform
exhaustive sampling of a large corpus of chemical systems and their
respective conformers, which leads to a 'universal' machine-learned
potential energy surface (UML-PES), also referred to as universal
machine-learned energy functions in the literature\cite{batatia2023foundation,yuan2025foundation,levine2025open,wadell2025foundation}. The
development of such UML-PES is appealing as it promises the
possibility of simulating multiple chemical systems at a low
computational cost. However, it is worth noting that UML-PES are
typically constructed based on DFT calculations, which limits the
accuracy of the model, in particular, when compared to experimental
measurements\cite{mannan2025evaluating}. In addition, many UML-PES are
constructed focusing on limited sets of organic molecules. Examples of
those general databases are those from the ANI
family\cite{smith2017anidata,smith2020ani,devereux2020extending},
QM7-X\cite{hoja2021qm7}, GEOM\cite{axelrod2022geom}, or
QMugs.\cite{isert2022qmugs} A complete description of the available
datasets for molecular simulations can be found in recent
reviews\cite{kriz2023imbalance,ullah2024molecular,kulichenko2024data}.
Complementary, community efforts for collecting existing datasets in a
centralized fashion have led to initiatives like ColabFit
exchange\cite{vita2023colabfit}. It should be noted that the mentioned
databases can be extended by adding new chemical compounds and/or
conformations; in such cases, the same level of theory and software
used to construct the initial database must be applied to ensure
compatibility.\\

\noindent
Despite the promise of UML-PES, they (still) do not provide the
required accuracy or description of certain interactions required for
specific applications. Therefore, it is common to create dedicated
databases either for fine-tuning of UML-PES (see below) or for
particular chemical systems. In this case of constructing dedicated
databases, it is necessary to strike a balance between data quality
and quantity. This concerns primarily the computational cost for data
generation which requires familiarity with the advantages and
shortcomings of the various electronic structure
methods.\cite{kulichenko2024data} The initial sampling for the
construction of a ML-PES can be performed using methods such as
\textit{ab initio} molecular dynamics\cite{behler2021four}, normal
mode sampling/scanning\cite{smith2017ani,MM.asparagus:2025}, virtual
reality
sampling\cite{o2018sampling,amabilino2019training,amabilino2020training,chu2022exploring},
sampling based on atoms-in-molecule (AMONS) fragments
\cite{lilienfeld2020slatm}, Diffusion
Monte-Carlo\cite{annarelli2024brief,conte2020full}, or enhanced
sampling methods (e.g. umbrella sampling\cite{torrie1977nonphysical}
or
metadynamics\cite{barducci2011metadynamics,herr2018meta,yoo2021metasampl}). In
addition, combinations of these methods can be used and are
particularly important for reactive
systems.\cite{brezina2023reducing,lee2025automated}\\

\begin{figure}
    \centering
    \includegraphics[width=\linewidth]{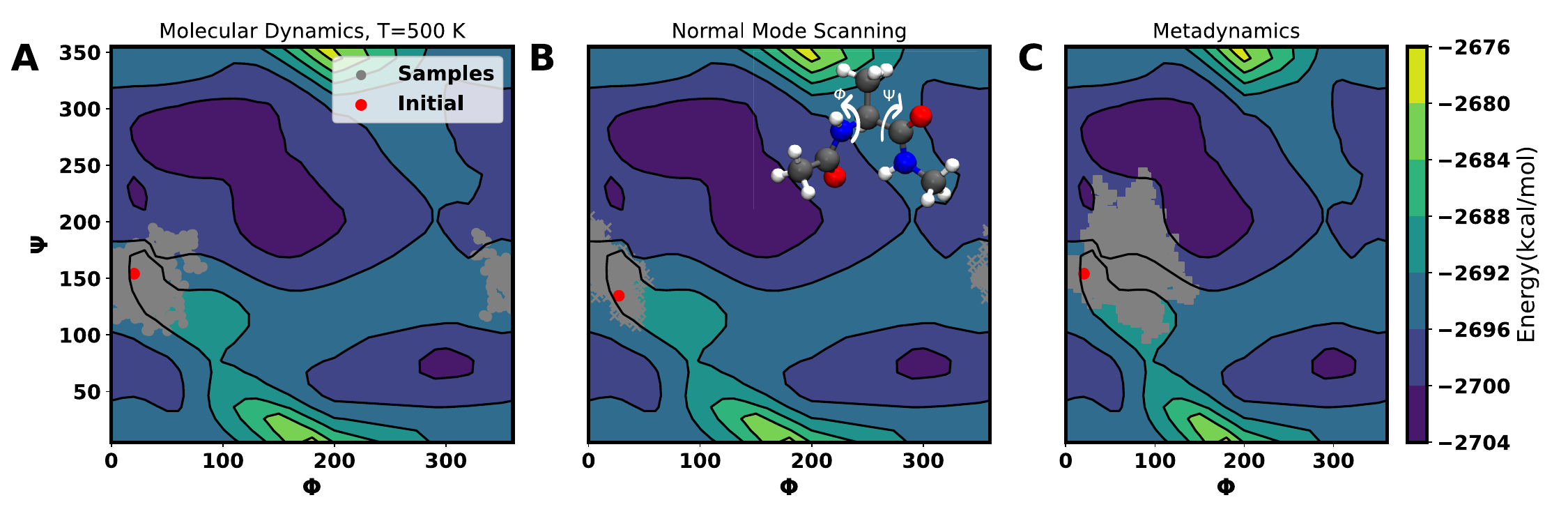}
    \caption{\textbf{Effect of different sampling methods for the
        dialanine peptide.} Panel A shows sampling by molecular
      dynamics at 500 K. Panel B corresponds to normal mode
      scanning. Panel C corresponds to metadynamics sampling at 500 K
      over the angles $\phi$ and $\psi$. For details about the
      calculation setup, look at the main text. In all cases, the
      calculations were performed at MP2/6-31G** using
      ORCA\cite{ORCA5}}
    \label{fig:sampling_diala}
\end{figure}

\noindent
As an example for the construction of a specific database, Figure
\ref{fig:sampling_diala} shows the conformations of dialanine peptide
visited by three different sampling methods. Figure
\ref{fig:sampling_diala}A displays the results from using very short
MD simulations (10 ps) at 500 K with a time step of 1 fs to generate a
total of 1000 samples. The results indicate that the generated
conformations are in the neighbourhood of the initial structure. This
is a natural consequence of how the MD method works. Therefore, it
should be noted that if the system of interest contains multiple valid
isomers, each must be simulated for a sufficiently long time scale. In
addition, it must be considered that the simulation of each system
should be done at a higher temperature than the planned used to avoid
the model entering an extrapolation regime where such NN-based models
usually become unstable and unreliable.\cite{unke2021machine}\\

\noindent
Figure \ref{fig:sampling_diala}B shows the results of using normal
mode scanning\cite{MM.asparagus:2025} considering the correlation of
two normal modes, excluding frequencies smaller than 100 cm$^{-1}$,
using an energy step of 0.05 eV, and a limit of 1 eV. An advantage of
using normal mode scanning with respect to molecular dynamics
simulation for generating the initial training dataset is the
efficient generation of uncorrelated samples\cite{MM.rev:2023};
however, those are centered near the equilibrium structure of the
initial molecule. Therefore, alternatives such as the combination of
multiple normal modes in normal mode scanning or a combination of
sampling strategies are necessary to obtain larger coverage of
conformational space.\\

\noindent
Finally, Figure \ref{fig:sampling_diala}C shows the results of
metadynamics sampling at 500 K over the angles of the dipeptide. It is
clear that the conformational space explored by this sampling method
is the largest of the methods previously described. However, the use
of these simulation methodologies is not straightforward as they
require knowledge of the collective variable to be sampled, in
particular, for reactive processes\cite{bussi2020using}, although it
is of less relevance for conformational
sampling\cite{herr2018meta,yoo2021metasampl}. \\

\noindent
In summary, it may be useful to search databases for existing data to
be reused and adapted for a particular project before generating new
training data. Often, the refinement of a model with a few high-level
calculations using transfer learning\cite{kaser2022fad},
morphing\cite{MM.morphing:2024}, or meta learning
strategies\cite{allen2024learning} is less computationally expensive,
and can achieve the required accuracy for a ML-PES for MD
simulations. Recently, the transferability between
models has been evaluated\cite{niblett2025transferability}. Of course, the accuracy of a trained model is directly
affected by the quality, level of theory, quantity and diversity of
the training data.\\

\noindent
Before training an ML model, the generated data must be cleaned to
avoid defective geometries, duplicate points, or molecules with
multireference character (i.e. corresponding to more than one
electronic state). The use of raw data is not advisable, as it is very
unlikely to obtain meaningful predictions from
it.\cite{medina2021rationality} It has been estimated that databases
commonly contain $\sim$ 10\% of incorrect data.\cite{artrith2021best}
Data curation is not an easy task, although some recommendations have
been put forward.\cite{chicco2022eleven} First, the data must be
scanned for the removal of duplicate points or inconsistent geometries
(i.e. points for which values of energy or force do not correspond to
the molecular geometry). Unfortunately, this may need to be done
manually, depending on the specific problem at hand, because
unphysical configurations can help models to make better predictions
by providing the ML model with information about high-energy
configurations.\cite{kulichenko2024data} In addition, recently, it was
shown that the addition of non-equilibrium structures can help in
predictions on chemical space.\cite{vazquez2025augmenting} A second
aspect to consider is the removal of outliers. Determining whether or
not a data point is an outlier is not an easy task. This step may
require complementary analysis on the dataset. Recently, through the
use of uncertainty, it was found that outliers in reactive potential
energy surfaces correspond to conformations with multireference
effects.\cite{MM.uq:2025} \\

\noindent
Multireference electronic effects emerge during sampling at high
temperatures, which leads to the generation of distorted geometries,
or when studying reactive processes involving bond breaking. These
effects are a consequence of the Born-Oppenheimer (BO) approximation,
which leverages the fact that nuclei are $\sim 2000$ times heavier
than electrons allowing the separation of electronic and nuclear
degrees of freedom and the appearance of electronic
levels\cite{gonzalez2020quantum}. Nevertheless, the BO approximation
fails when the atomic distances change and multiple solutions to the
electronic Schr\"odinger equation are available, as it is common in
photochemistry and high-energy processes. Current ML-PES are not able
to describe multiple electronic states in a single model as most of
them are designed to operate in the electronic ground states. The
reasoning behind such constraints is that the study of chemistry in
excited electronic levels requires special
methodologies\cite{knepp2025excited,prlj2025best}, large amounts of
data, and advanced sampling techniques. A detailed discussion of such
approaches lies beyond the scope of this work; readers are referred to
Refs.~\cite{westermayr2020machine,muller2025machine} for comprehensive
overviews of current methodologies and best practices. Nevertheless,
it should be mentioned that when constructing a ML-PES, that does not
implicitly consider multiple electronic states, their presence leads
to inconsistencies in the training data, confusion in the model, and
discontinuities in the
fitting\cite{unke2021machine,kulichenko2024data}. Therefore, it is
necessary to remove samples that present multireference effects. A
possible method to do it is by following the methodology proposed by
Behler\cite{behler2021four} of comparing the error of multiple
independently trained models.\\

\noindent
Finally, the {\it trainability} of a given NN-architecture may depend
on the level of theory at which reference data was determined. For
example, it was found that a NN (here PhysNet was used) using the same
architecture and hyperparameters reached error levels differing by
almost one order of magnitude depending on whether the reference data
for HONO was computed at the MP2 or MRCI+Q levels of
theory.\cite{MM.error:2023} One possible explanation for this is that
multi-reference methods involve a truncation of the full configuration
interaction expansion which introduces ``noise'' in the data to which
the NN is susceptible during training. However, it is possible that
there are additional reasons for the dependence of the NN-training
depending on the source of the reference data. It is noted that for
CASPT2 reference data, similar observations were
made.\cite{MM.criegee:2023,MM.h2coo:2024,MM.h2coo:2025}\\

\noindent
\textbf{Recommendations:} Data sets need to be generated at
sufficiently high level of quantum chemical theory to be suitable for
the problem to be solved. Clean data is mandatory for obtaining robust
and meaningful high/full-dimensional ML-PESs. This is particularly
important when advanced quantum chemical methods, including CASPT2- or
MRCI are used.\\

\section{Validation and Improvement of ML-PESs}
Validation is essential in developing and deploying ML-PESs. This
ensures that the model accurately represents the PES across diverse
regions of configurational space accessible to the system. Unlike
traditional EEFs, which rely on predefined functional forms and
physical constraints, ML-PESs are inherently data-driven and can be
prone to effects such as overfitting and instability in regions where
training data are sparse. A robust validation process not only
assesses the accuracy/the fitting error of the model on a test set but
also its stability and suitability for various downstream tasks such
as MD simulations. The examples presented in sections 5 and 6
illustrate explicitly the points discussed below. \\

\subsection{Characterisation of a Trained Model}
Conventional quantification of the quality of a trained ML-PES
consists of reporting statistical measures on a test set. It is
important to notice that the error in a training dataset is not a
reliable metric for characterising an ML-PES, as this quantity can be
reduced infinitely during the optimisation procedure. Therefore, in
typical training setups, a randomly chosen $\sim 10$\% of the
available data is set aside for testing. For this hold-out of the data
set that the model has never ``seen'', quantities such as the ``root
mean squared'' or ``mean averaged'' errors (RMSE or MAE) are
determined; other example of quantities can be found in
Ref. \cite{bender2022evaluation}. It is necessary to mention that when
computing MAE for vectorial quantities (i.e. forces), this quantity is
not $E(3)$-equivariant depending on the coordinate system and has
biases on conformation aligned to the
axes\cite{wang2025design}. Therefore, RMSE is preferred for computing
the error of vectorial quantities.\cite{morrow2023validate}
Furthermore, it must be noted that the pure values of MAE or RMSE do
not suffice to characterise the statistical error of an ML-PES;
additional statistical quantities, such as the Pearson correlation
coefficient or the determination coefficient, should be calculated to
compare the predicted and reference values. On the other hand, other
simple analyses, such as residual plots (i.e., the energy range of the
training data vs. error) or the distribution of errors, can help to
identify regions that require further sampling, outliers, or
non-converged samples. \\

\noindent
Complementary to the statistical validation of an ML-PES, a physical
characterisation is required. An ML-PES can only be considered
validated if it reproduces experimental or numerical observables. A
simple, yet very effective, way to determine the quality of the
obtained ML-PES is to determine the optimised structures and normal
mode frequencies, and compare those with electronic structure
calculations at the level of theory that was used for reference data
generation. Complementary, the ML-PES must be monitored for
stability. It is known that ML-PES, when running dynamical
simulations, can lead to unphysical states such as bond-breaking when
not trained for it. Therefore, the stability of the trajectory can be
monitored by standard procedures (i.e. energy
conservation\cite{tuckerman1992reversible}) or more elaborate methods
(i.e. RBF or bond distance
changes\cite{fu2023forces,niblett2025transferability}). It is worth
mentioning that for reactive simulations, the estimation of the
minimum energy path\cite{marcus:1966,fukui:1970} and minimum dynamic
path\cite{unke2019sampling} are a requirement to ensure the stability
of the ML-PES. The determination of the stability of an ML-PES is
critical for large simulations to avoid artifacts or undesirable
problems.\\

\noindent
A third level of sophistication is afforded by running explicit
simulation and to determine experimental observables for comparison
and
validation\cite{morrow2023validate,ranasinghe2025basic,MM.tl:2022,MM.tl:2025}. Such
tests not only validate the trained ML-PES {\it per se} but also the
level of theory that was used for data generation. All computable
properties for gas- and condensed phase systems can be used for such a
comparison, including infrared spectra, reaction rates, diffusion
coefficients, thermal expansion coefficients, to name a few.\\

\noindent
\textbf{Recommendations:} For validation of a ML-PES the usual
statistical measures (RMSE and/or MAE) for energies, forces and
potentially other/additional properties need to be determined. For
test data, graphical representations of properties from the ML-PES
versus the reference quantum chemical method are required. Validation
against known experimental or numerical quantities is highly
recommended.\\

\subsection{Robustness of and Improving the Trained Models}
When conceiving and training a ML-PES, the initial sampling procedure
may have, inadvertently, excluded specific configurations. Such
undersampled regions can manifest themselves as "holes" which are
configurations with lower energy than the global minimum. For
detecting such regions, diffusion Monte-Carlo (DMC) simulations, which
use random walkers to explore conformational space, are highly
valuable. The details of DMC lie beyond the scope of this work, and
interested readers are referred to the relevant literature for further
information\cite{anderson1975random,kosztin1996introduction,conte2020full,MM.transfer:2022,annarelli2024brief}. Notice
that exhaustive sampling of configurational space using DMC can become
computationally expensive. Therefore, it is often used only for
constructing high-quality ML-PESs.\\

\noindent
Complementary to the search for holes, a necessary aspect for
constructing more accurate and robust ML-PES is quantifying the
uncertainty associated with the predictions. Currently, several
methodologies are available for making such evaluations. Recent
benchmarks for non-reactive\cite{bombarelli:2023} and
reactive\cite{MM.uq:2025} ML-PES have been published. The most common
procedure for determining the uncertainty in prediction is the
ensemble method, where several ($\sim 3-5$) neural networks are
independently trained, usually at the same time. Then, the energy,
forces or other quantities are predicted, and their mean and standard
deviation are determined. Regions that are undersampled are identified
by large disagreement (i.e., large variance) in the predictions of
independently trained models\cite{gastegger2020molecular} which is
known as ``query-by-committee''.\cite{seung1992query} Such techniques
typically require substantial amounts of computational
resources. Consequently, alternative methodologies, such as Gaussian
mixture models\cite{zhu2023fast} (GMM), deep evidential
regression\cite{vazquez2022uncertainty,MM.uq:2025} (DER), shallow
ensembles\cite{kellner2024uncertainty}, conformal
prediction\cite{hu2022robust}, etc. have been conceived. The ultimate
goal to apply any of these techniques is to have a starting point for
active learning.\cite {settles2012active} Here, new samples are added
to the initial database to refine the ML-PES and to improve its
accuracy over the course of active learning iterations. \\

\noindent
\textbf{Recommendations:} Trained ML-PESs must be checked for holes
and discontinuities (e.g. by using DMC) to avoid, for example,
sampling nonphysical structures during production simulations. The
extrapolation capabilities outside the coverage of the training set
need to be assessed. It is recommended to consider reducing the
dataset sizes for trained models to avoid imbalances in the
distribution of training samples. A future question is whether there are additional,
more meaningful, and robust measures of performance that can be developed and applied. Uncertainty in the predictions should be quantified, see
"Challenges".\\

\section{MD Simulations Using ML-PESs}
ML-PESs can be used to characterize chemical systems ranging from gas
phase molecules to condensed phase or solid-state
systems.\cite{zahn.mlff:2024,ml.cond:2024} Similar to mixed quantum
mechanical/molecular mechanics (QM/MM) approaches, mixed ML/MM
representations are also possible.\cite{MM.fad:2022,MM.eutectic:2022}
For such situations various embedding schemes have been developed to
describe the interaction potential between the QM (ML-PES represented)
atoms of the chemical subsystem and the MM atoms.  \\

\noindent
Training of ML-PES for complete chemical systems can be challenging,
as it becomes more difficult to determine whether the model has
adequately covered all crucial parts of conformational
space. Additionally, training a model on larger chemical systems
increases computational costs. Therefore, as mentioned before, the use
of published UML-PES or semi-UML-PES trained on a large amount of
chemical compounds is an appealing
option.\cite{ml.gnome:2023,ml.guide:2025} To construct ML-PESs for
condensed phase systems, the initial sampling can be carried out by
using EEFs followed by DFT calculation of sampled
structures.\cite{madsen:2022il,abel:2022le, zahn.mlff:2024} Such a
ML-PES can be further refined by using an active learning
algorithm. Together, the sum of initial sampling, training, and active
learning steps is implemented in the VASP suite\cite{kresse1996vasp}
or in the python package AUTOFORCE\cite{kim:2021autoforce} allowing
for a ``on-the-fly machine-learning algorithm''. \\

\noindent
The ML/MM approach has the advantage of combining computationally
computationally efficient EEFs with ML-PES for smaller molecular
subsystems.  The smaller size of the system treated by the ML-PES does
not just decrease the computational costs, but also allows quantum
mechanical reference calculation at a higher level of theory.\cite{
  rufamlmm:2020} The ML/MM interaction potential can be either treated
through Coulomb's law between fixed charges assigned to both MM and ML
atoms or between fixed MM atom charges and fluctuating charges of the
ML system such as available from
PhysNet.\cite{MM.physnet:2019,MM.fad:2022} Such fluctuating ML point
charges and even higher multipoles can be determined by the ML system
configuration (mechanical embedding), but also from external
polarization through the MM atom charges (electrostatic
embedding).\cite{lei:mlmm:2024,riniker:2025,semalak:2025,neqoip:2025}
The embedding scheme affects the electrostatic contribution to the
non-bonding interaction potential between the ML and MM subsystems. On
the other hand, the remaining non-bonded interactions (van-der-Waals)
are still represented empirically by using a Lennard-Jones (LJ)
potential with pre-defined atom-type specific LJ parameters and
corresponding combination rules.\\

\subsection{Using ML-PESs in established MD Software}
With the development of UML-PES and semi-UML-PES models, as well as
easier access to self-trained ML-PES models, the number of their
applications in dynamics simulation or methods with repeated potential
evaluations have increased. For that purpose, interfaces were
implemented between ML-PES models and established atomic calculation
programs such as ASE\cite{larsen:2017}, LAMMPS,\cite{lammps:2022}
OpenMM,\cite{openmm8:2024} Amber\cite{ambertools:2023},
GROMACS\cite{gromacsNNPOT}, or CHARMM\cite{MM.charmm:2024} via its
pyCHARMM interface.\cite{buckner:pycharmm:2023,MM.pycharmm:2023}\\

\noindent
To the best of the authors' knowledge, most ML-PES model
implementations have an interface to an ASE calculator
class\cite{larsen:2017}, which allows to combine the ML-PES models
with the methods implemented in ASE. As ASE is implemented in Python
computational performance is slower than implementations in, e.g., C++
or Fortran. This can become problematic for MD simulations of large
molecules such as proteins. Alternatives for high-performance
simulations on CPUs or GPUs can be done by using packages such as
Schnetpack\cite{schnetpack2.0:2023} that provide interfaces of the
implemented models in it to the LAMMPS software. Alternatively, the
plugin OpenMM-ML connects several model architectures such as
ANI,\cite{torchani:2020} MACE,\cite{batatia2022mace,batatia:2025}
Nequip\cite{batzner2022nequip} or DeepMD\cite{wang2018deepmd} with the
OpenMM engine. Lastly, the Asparagus package provides interfaces
between implemented and trained ML-PES models (currently PhysNet and
PaiNN) with CHARMM and OpenMM.\cite{MM.asparagus:2025} \\

\noindent
\textbf{Recommendations:} A suitable choice for an MD program that
supports ML-PES models depends on the scientific question asked. For
example, a MD software is already in use for the simulation of a
chemical system but one wants to switch from EEFs to a ML-PES models
or, the opposite situation, a trained ML-PES model is already
available but it is only applicable with a certain or even just one MD
software.  There is no better way to recommend using a ML-PES or MD
software which is available and already tested in combination with
each other, except one is experienced to write an interface of its own
between both.  From the current perspective, such interfaces are
becoming more and more available with the recent updates of both MD
and ML-PES software.\\

\begin{figure}
    \centering
    \includegraphics[width=\textwidth]{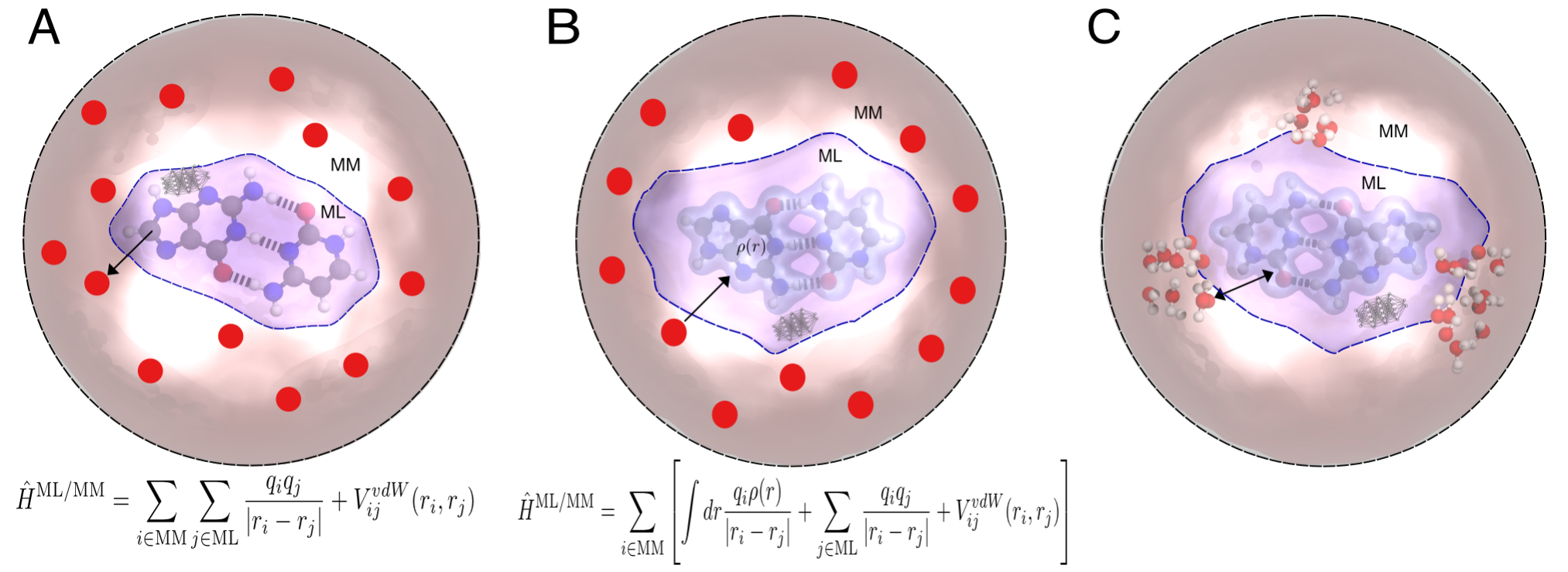}
    \caption{\textbf{Types of ML/MM embeddings.} The central region
      (blue) represents the subsystem described using machine-learning
      (ML) methodologies, surrounded by the molecular-mechanics (MM)
      environment (light red). Red spheres denote solvent molecules
      modelled as point charges. Panel A: mechanical embedding where
      the ML region interacts with the MM environment only by
      electrostatic and van der Waals interactions; the corresponding
      Hamiltonian for the ML–MM coupling is shown below. Panel B:
      electrostatic embedding for which the ML region is polarised by
      the MM point charges, as indicated by the electronic density
      interacting with the MM electrostatic field. The corresponding
      Hamiltonian is shown below. Panel C: schematic representation of
      more elaborate embeddings, such as polarizable or adaptive
      schemes. Notice in this case, the solvent could enter the ML
      region and/or the ML region also polarize the MM region.}
    \label{fig:embeddings}
\end{figure}

\subsection{Mechanical Embedding}
The conceptually simplest way to compute the potential of a ML/MM
system by separate evaluations of the ML-PES model and the empirical
energy function, and to describe the interaction potential between ML and
MM atoms by non-bonded electrostatic and vdW contribution using
empirical parameters.  This approach is referred to as ``mechanical
embedding'', see Figure \ref{fig:embeddings}A. The polarization of the
ML system by the electric field originating from MM atoms is
neglected. The electrostatic contribution to the interaction potential
can be between fixed MM atom-centered charges and static ML
atom-centered charges, fluctuating ML
charges,\cite{MM.fad:2022,MM.pycharmm:2023} or other forms of
sophisticated electrostatic models. The static LJ parameters in the
vdW terms are pre-defined for each atom type and optimized for a
specific form of electrostatic interaction (mostly fixed atom-centered
charges).\\

\noindent
Replacing the empirical non-bonded electrostatic interaction terms
leads to the question whether the vdW contribution with its LJ
parameter needs to be reparametrized. A typical example is the
improvement of a point charge-based representation of the
electrostatics in EEFs by multipolar representations for condensed
phase simulations.\cite{MM.mtp:2012,MM.mtp:2013,MM.mtp2:2013} Because
in typical parametrization protocols\cite{cgenff:2012} LJ parameters
are fitted in the final step to condensed phase properties of the pure
liquid, they directly depend on the charge model that was used in
these simulations. Hence, when introducing more advanced
representations for the electrostatics, such as distributed charge
models,\cite{MM.dcm:2014,MM.dcm:2017,MM.dcm:2022,MM.kmdcm:2024} the LJ
parameters require further refinement.\\

\noindent
An alternative procedure consists of considering cluster models for
pure substances for which electronic structure calculations at
sufficiently high level of theory. Interaction energies from such
calculations can then serve as a proxy to improve the non-bonded
interactions and to test them in condensed phase simulations. Such a
"cluster-based approach" has been recently employed for N$_2$O in
Argon or SF$_6$ or for eutectic
mixtures.\cite{MM.eutectic:2022,MM.n2o:2024,MM.cluster:2025}\\

\noindent
\textbf{Recommendations:} When the scope of the project extends, e.g.,
from gas phase simulations to condensed phase with a solute
represented by a ML-PES model and a solvent with empirical force
fields, mechanical embedding scheme is the simplest to apply if an
empirical non-bonding parameter set is available for the solute within
the EEF. Experience from QM/MM simulations showed that the equivalent
mechanical embedding scheme already yield good agreements with respect
of the structure changes between gas and condensed phase. More
advanced embedding schemes become essential if external polarization
of the ML atom system by the solvent is requested, e.g. for predicting
infrared spectra.\\

\subsection{Advanced Embedding Schemes}
``Electrostatic embedding'' is the current gold standard for QM/MM
simulations although arguments have been put forward that
``polarizable embedding'' is even
superior.\cite{rothlisberger:qmmm:2015,mennucci:qmmm:2020} The ML
charge distribution in ML/MM calculations also depends on the external
polarization originating from the electric field of the MM atom
charges (Figure \ref{fig:embeddings}B). For polarizable embedding a
mutual polarization between the ML and MM charge distributions is
possible if the MM system is represented by an empirical polarizable
energy function (Figure \ref{fig:embeddings}C).\\

\noindent
To mimic electrostatic embedding in ML/MM calculations, the ML-PES
models responds to an external polarization either by a {\it post hoc}
polarization correction scheme or by explicitly including the MM
environment in the input of the ML-PES model. The polarization scheme
has the advantage that the ML-PES can still be trained with reference
data of the isolated molecular system in vacuum and the polarization
schemes can be adopted from established empirical polarizable force
fields such as the Thole model used in
AMOEBA.\cite{thole:1983,amoeba:2010,zinovjev2023electrostatic,inizan2023scalable}
Providing the MM atom charge and position information as input for the
ML-PES model allows to reproduce full electrostatic embedded QM/MM
reference calculation. Additional sampling of the MM environment in
addition to sampling the ML configurational space is required, but it
can yield excellent accuracy in reproducing the external polarization
of the ML charge
distribution.\cite{gastegger2021mlsolv,riniker:2025,gao:mlmm:2025,neqoip:2025}
An extensive review of embedding schemes for ML/MM simulations is
provided in Ref. \citenum{semelak:mlmm:2025}.\\

\noindent
Even though the electrostatic representation of the ML atoms is
improved, the empirical LJ-parameters still need to be readjusted once
the electrostatic model is modified. However, the combination of
empirical ML-MM interaction potential with the improved response of
the ML charge distribution with respect to the MM environment already
leads to increased accuracy for properties such as IR
spectra.\cite{gastegger2021mlsolv}\\

\noindent
\textbf{Recommendations:} Using electrostatic embedding schemes for
hybrid ML/MM simulation can be either achieved by applying {\it post
  hoc} polarization corrections on top of the ML-PES model trained on
a chemical system in vacuum or by using a ML-PES model that supports
external polarization, which already limits the selection choices for
ML-PES models. Aiming towards electrostatic embedded ML/MM simulation
preparation is above the scope of this work.\\

\section{Application to a Non-reactive System: Tripeptides}
The final part of this work presents two examples of constructing an
ML-PES. First, a tripeptide system would be described from the data
generation step to characterization and use for molecular dynamics
simulations. First, the data generation and model training will be
described, followed by the presentation of results, including
comparisons with EEFs.

\subsection{Data Generation and Model Training}
The initial models for the tripeptide Alanine-Lysine-Alanine (AKA)
were generated and trained following standard procedures. First,
molecular dynamics simulations in the gas phase were carried out using
CHARMM\cite{charmm:2009,MM.charmm:2024} code and the CGenFF force
field.\cite{cgenff:2012} In addition to the harmonic bonds involving
hydrogen atoms, all X-H (X: C, N or O) bonds were replaced by Morse
oscillators with reduced force constant to allow larger-amplitude
motions. The reasoning behind this choice is to provide a better and
more robust description of bonds involving H-atoms for the training of
the ML-PES.  This was done in anticipation of training a NN-based
energy function to provide better and more robust extrapolation
capabilities in particular for bonds involving H-atoms.  The Morse
parameters were $D_e = 40.0$ kcal/mol, $\beta = 1.0$ \AA\/$^{-1}$, and
$r_e = 1.5$ \AA\/ which describe considerably softer bonds than
physical (N-H, C-H, and O-H) bonds. The time step in all simulations
was $\Delta t = 0.1$ fs and samples were obtained every 25 ps.\\

\begin{figure}[h!]
\begin{center}
\includegraphics[width=0.9\textwidth]{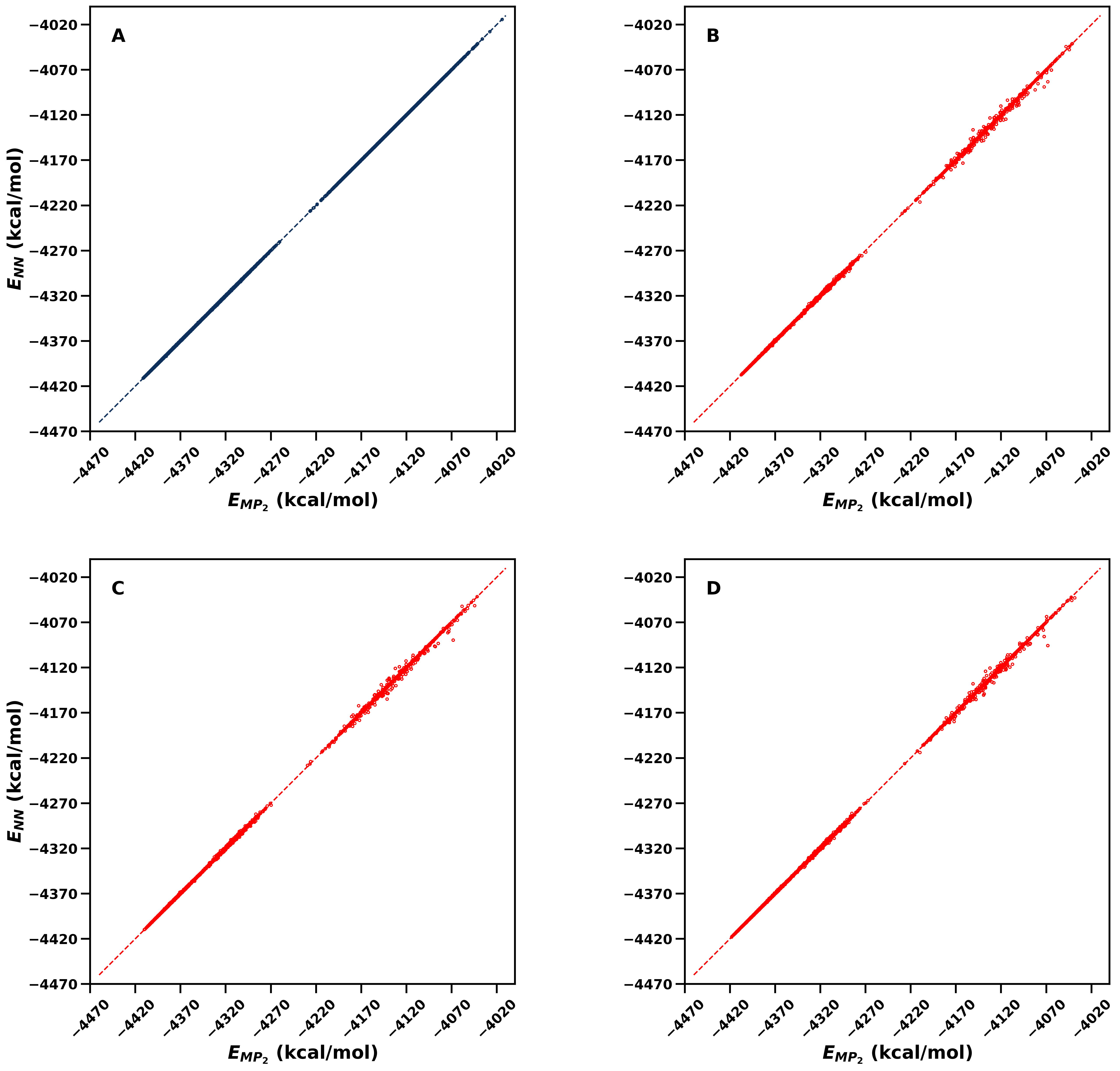}
\caption{\textbf{General performance of the ML-PES for AKA tripeptide}
  Correlation plot of train (A) and test (B) set of SETG5 reference
  energies and predicted NN energies. RMSE(E) and MAE(E) for the
  training and test sets of SETG5: For the training set, RMSE(E) is
  0.14 and MAE(E) 0.1 kcal/mol; for the test set, RMSE(E) is 1.31 and
  MAE(E) is 0.5 kcal/mol. Correlation plot of test set of SETS1 (C)
  and SETS2 (D) reference energies and predicted NN energies. RMSE(E)
  and MAE(E) for the test set of SETS1 are 1.32 and 0.5 kcal/mol,
  respectively. For the SETS2 test set, RMSE(E) is 1.41 and MAE(E) is
  0.52 kcal/mol.}
\label{fig:corr_set5}
\end{center}
\end{figure}

\noindent
From the initial simulation, 10001 structures were obtained at the
MP2/6-31G level of theory with the MOLPRO \cite{molpro:2011} suite and
used to train two independent ML-PESs based on the PhysNet
architecture. This was followed by several rounds of adaptive
sampling. In each round of adaptive sampling, several structures
(between 400 and 1250) were extracted and used to generate new samples
for which energies and forces were obtained.  Next, several rounds of
adaptive sampling followed. In each round between 400 and 1250
structures were extracted and reference energies and forces were
determined again at the MP2/6-31G level of theory.  Finally, six
structures were visually chosen as the most likely candidates for
metastable intermediates. From the selected structures, an MD
simulation was performed to generate a set of 1250 samples, which were
then added to the training dataset. The final dataset, labelled SETG5,
consisted of a total of 17649 structures. The performance of the
trained model on the test structures is reported in Figures
\ref{fig:corr_set5}A and B. \\

\subsection{Training including AKA Structures in Solution}
As a next step, additional structures obtained in explicit solvent
were incorporated into the datasets for training two new models,
starting from SETG5. In this case, all ML/MD simulations were
performed with CHARMM software and its pyCHARMM
interface.\cite{MM.charmm:2024,buckner:pycharmm:2023} For the ML/MD
simulations, the PhysNet model based on SETG5 was used to
describe the ML part, while for the empirical part CGenFF was
employed. The simulations were set up using a time step of $\Delta t =
1$ fs. Snapshots were collected every 500 ps with $\Delta t = 0.25$ fs
time step. The first dataset comprised the structures SETG5 and 100
new configurations generated from ML/MD simulations with CGenFF
charges in the condensed phase. In the second round of addition, 1000
structures from ML/MD simulations, where CGenFF charges were used, and
200 structures from the ML/MD simulations, in which fluctuating
charges were assigned, were considered. The RMSE and MAE evaluated the
performance of the two trained models with the additional solvated
conformation for both energy and force, see Figure
\ref{fig:corr_set5}. For SETS1, the RMSE for energy predictions was
1.32 kcal/mol (MAE: 0.50 kcal/mol) on its test set. Complementary, the
RMSE value of forces in the test set was 2.0 (kcal/mol)$\cdot \rm
\AA^{-1}$ (MAE:0.86 (kcal/mol)$\cdot \rm \AA^{-1}$). In comparison,
SETS2 showed slightly larger RMSE values, with 1.41 kcal/mol (MAE:
0.52 kcal/mol) for energy and 2.1 (kcal/mol)$\cdot \rm \AA^{-1}$
(MAE:0.87 (kcal/mol)$\cdot \rm \AA^{-1}$) for forces in the test
set.\\

\subsection{Results}
Using the ML-PES, ML/MD simulations were conducted with CHARMM code in
both gas and condensed phases. The simulations were performed with a
time step of $\Delta t = 1$ fs, for a total duration of 1 ns in the
gas phase and for 1.5 ns in solution, respectively. ML/MD simulations
were performed applying ML-PES (SETG5 model) at $T = 300$ K and
SHAKE\cite{shake} algorithm was used to constrain all bonds involving
hydrogen atoms. In solution, the system consists of a cubic water box
with dimensions of (41$^3$) {\AA}$^3$. For water, TIP3P model in
CGenFF energy function was used. The SHAKE algorithm was only applied
to water molecules. After heating and equilibration parts, an $NPT$
simulation was carried out at 300 K using the leap algorithm. ML-PES
was employed for AKA structure throughout the simulation. The charges
predicted from the PhysNet model were calculated for every 2.5 ps, and
their median was calculated to be used in the analysis.\\

\begin{figure}[h!]
\begin{center}
\includegraphics[width=1\textwidth]{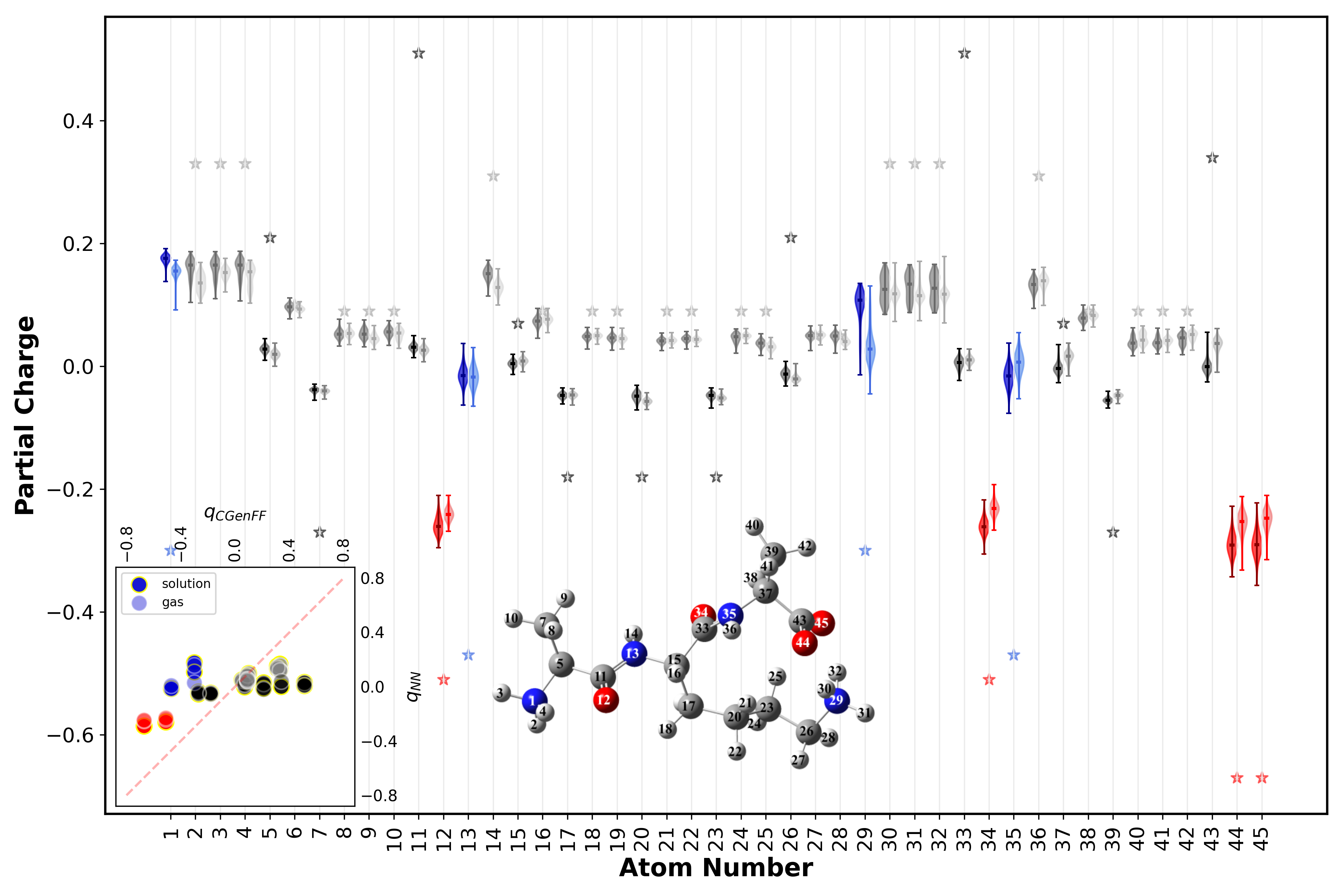}
\caption{\textbf{Partial charge distributions of the tripeptide AKA
    from ML/MD simulations.} Partial charge distributions of
  tripeptide AKA conformations obtained from ML/MD simulations in the
  gas phase (1 ns, 4000 structures, light colors) and condensed phase
  (1.5 ns, 6000 structures, dark colors). Structures were sampled
  every 2.5 ps. Distributions are color-coded by atom type: nitrogen
  (blue), oxygen (red), carbon (dark gray), and hydrogen (silver).For
  each distribution, the central line shows the median. The $x$ axis
  corresponds to the number of each atom in the 3D representation
  inside the plot. For comparison, CGenFF reference charges are
  indicated by stars in the corresponding colors. Inset: The
  correlation plot of fluctuating charges from ML/MD simulations and
  CGenFF charges in both gas and condensed phase.}
\label{fig:charges}
\end{center}
\end{figure}

\noindent
Figure \ref{fig:charges} shows the distribution of partial charges for
atoms in the tripeptide during the ML/MD simulations. Structures were
sampled every 2.5 ps over trajectories of 1 ns in the gas phase and
1.5 ns in solution. The fluctuating charge distributions are shown as
violin plots for gas-phase (light colors, right) and solution (dark
colors, left) using the respective color-code for each atom type.
From the results in Figure \ref{fig:charges}, nitrogen and oxygen
atoms feature a broader range of charge values in comparison to H- and
C-atom, with the exception of the hydrogen atoms bonded to O and N
atoms. The range of charges for N-atoms in the ML-PES covers
[-0.07,+0.19]$e$, and for oxygens it is [--0.35,--0.21]$e$. This is
remarkably different from the values observed for the same atoms when
using the CGenFF energy function, with ranges of [-0.47,--0.33]$e$ for
N and [-0.67,-0.51]$e$ for O. For example, in the terminal N-atom
(atom number 1 in Figure \ref{fig:charges}, the atomic charge
fluctuates between 0.15$e$ and 0.19$e$. Interaction between the atoms
in the molecule or with the environment, when in solution, leads to
changes in the fluctuating charges. To illustrate this influence,
consider the hydrogen atoms H2, H3, and H4 bonded to nitrogen N1. In
the gas phase, these atoms interact with the oxygen atom O34 due to
rotation around N1. H2 and H4 form hydrogen bonds with O34, which
strongly influence their partial charges. This effect is evident in
Figure \ref{fig:charges}, where the charge distributions of H2 and H4
spread over a broader range than H3. A further example is related to
the salt bridge formed in both gas and condensed phases. In peptides,
those are formed between oppositely charged residues; here, they occur
between R-COO$^{-}$ (C43-O44-O45) and H$_3$N$^{+}$-R (cationic group
of lysine residue, N29-H30-H31). In a salt bridge, the atoms are
close, influencing and being influenced by electrostatic attraction
\cite{salt.bridges}, which leads to changes in partial charges during
ML/MD simulations. Evidence of the effect of this interaction is found
when examining the charge distributions for the atoms involved. \\

\begin{figure}[h!]
\begin{center}
\includegraphics[width=0.9\textwidth]{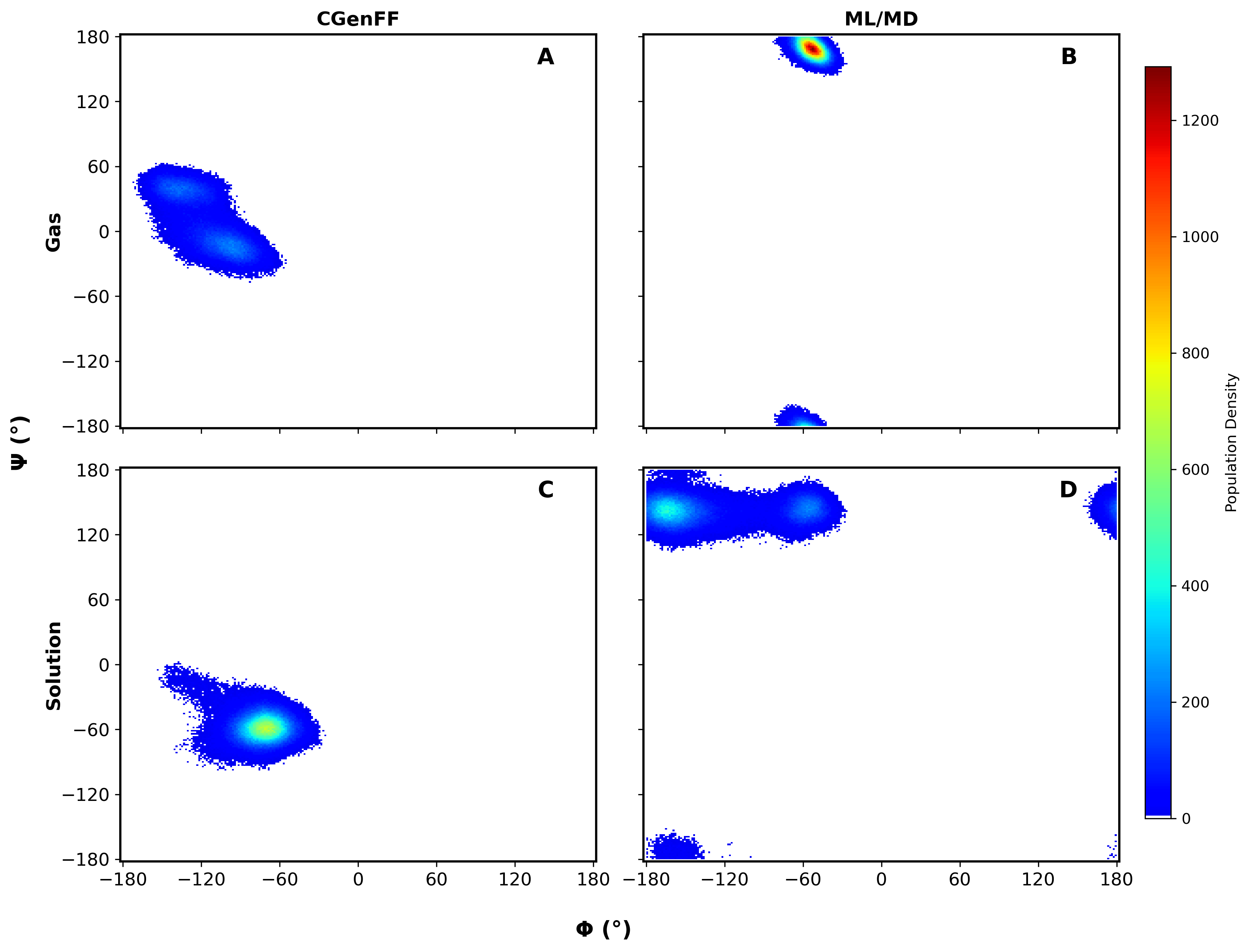}
\caption{\textbf{Ramachandran plot of AKA from CGenFF and ML-PES.}
  Panels A and C are from simulations using the CGenFF energy function
  in the gas phase and in solution, respectively. Panels B and D use
  the ML-PES (PhysNet) trained on SETG5.}
\label{fig:rama}
\end{center}
\end{figure}

\noindent
To characterize the conformations of AKA, Ramachandran plots were
constructed.  Figures \ref{fig:rama}A/C report results from
simulations using CGenFF in the gas phase and in solution, whereas
panels B and D are the same but using the ML-PES trained on the
dataset SETG5. When comparing the gas phase simulations (panels A and
B), it is found that using CGenFF leads to a larger number of
configurations. In particular, the fraction of $\alpha$-helical
structures is larger. This is consistent with previous observations
that without applying particular corrections empirical energy
functions for protein simulations are typically "too
helical".\cite{best:2008} On the other hand, using the ML-PES leads to
a propensity of $\beta-$sheets. Including the solvent leads to
increased flexibility in the backbone of AKA. \\

\noindent
Unexpected conformations were formed at $\phi / \psi-$ angles around
$[175^\circ,150^\circ]$ and $[-160^\circ,-170^\circ]$. Additionally, it was found that
ML/MD exhibits denser and more restricted conformations compared to
using CGenFF in both phases. Eker \textit{et. al.} studied various
tripeptides (AXA) with range of different analyses. Their findings
regarding cationic AKA indicate that the structure favors a PPII
(polyproline II) conformation.\cite{eker.axa} The analysis, conducted
through a combination of Raman and VCD spectra, was used to determine
the dihedral angles. The obtained dihedral angles ranges are $\phi \in
[-55 \pm 10]^\circ$ and $\psi \in [150 \pm 20]$.\cite{eker.axa} When
these angles were compared with data from the simulations, especially
for the solution phase, they showed similar ranges. \\

\begin{figure}[h]
\begin{center}
\includegraphics[width=0.9\textwidth]{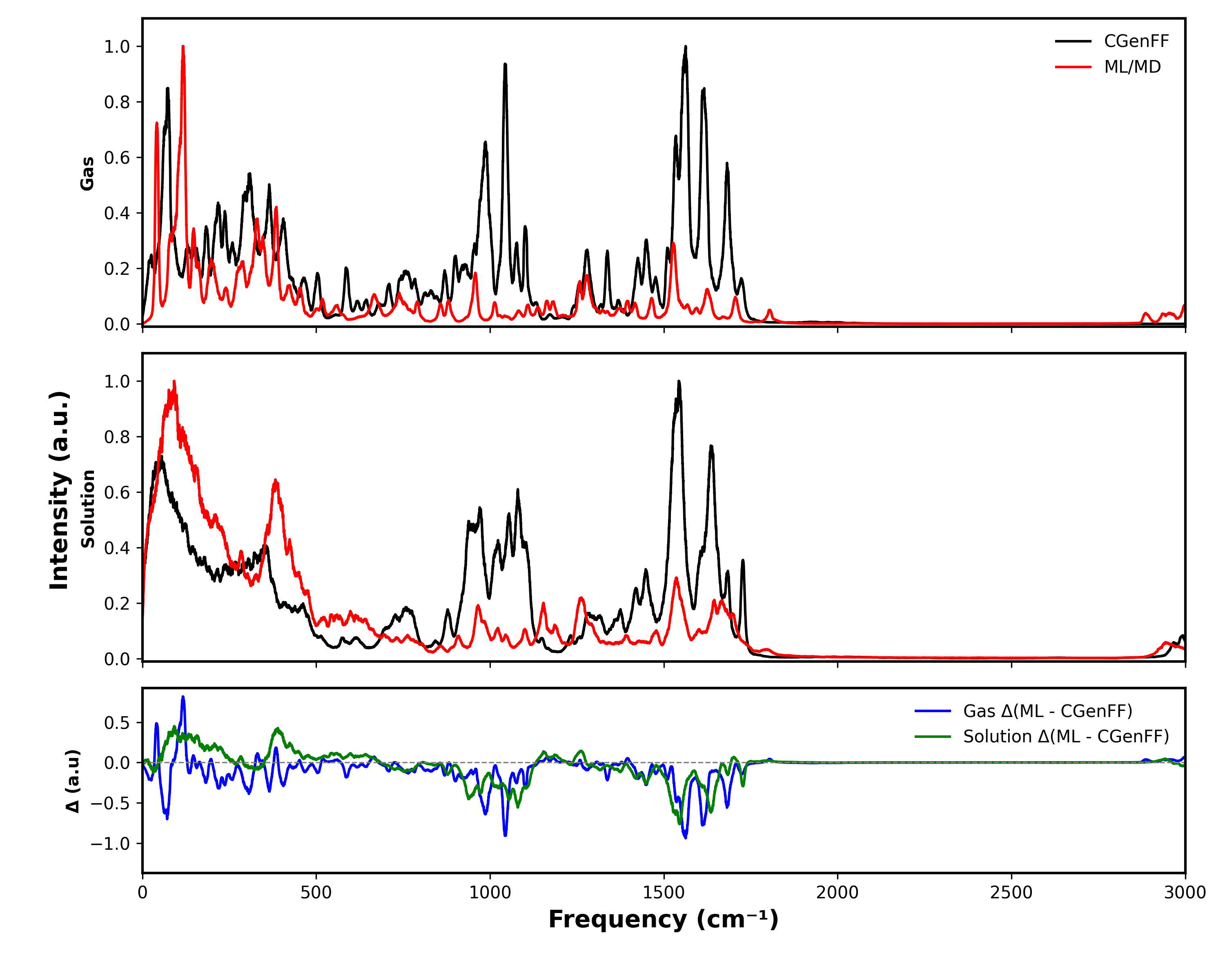}
\caption{\textbf{Infrared spectra of AKA from CGenFF and ML-PES
    simulations.} IR spectra obtained from simulations using CGenFF
  (black) and the ML-PES (red) in the gas phase (top panel) and in
  solution (centre panel). The difference between the gas (blue) and
  solution (green) spectra using the two energy functions is shown in
  the bottom panel.}
\label{fig:ir_ala_all}
\end{center}
\end{figure}

\noindent
Complementary to the previous analysis, the Infrared (IR) spectra was
calculated. This quantity provides information about the vibrations of
the molecule that reflect the bond types and the environment, with the
additional advantage of being easily obtained experimentally. IR
spectra is sensitive to changes between gas-phase and solution.  In
Figure \ref{fig:ir_ala_all}, IR spectra in both gas (top panel) and
solution (middle panel) phases were analyzed by performing MD
simulations using CGenFF and the ML-PES. The amide I region (1600 --
1800 cm$^{-1}$), associated with C=O stretching, is essential for
analyzing the main vibrational bonds of peptides and plays a key role
in understanding distinguishing features such as secondary
structures.\cite{Barth_Zscherp_2002} In the gas phase, the peaks
between 1500 -- 1800 cm$^{-1}$ are clearly visible in both
simulations. However, around 1700 and 1725 cm$^{-1}$ a double peak is
observed in the simulations using CGenFF, whereas this pattern is
shifted in the ML/MD simulations. In the solution phase, peaks in a
similar range from the ML/MD simulations merge, resulting in a broader
spectral range. In this case, shifting the IR spectrum from the ML/MD
simulations towards the blue yields a closer match with the results
using the EEF.\\

\begin{figure}[h]
\begin{center}
\includegraphics[width=0.9\textwidth]{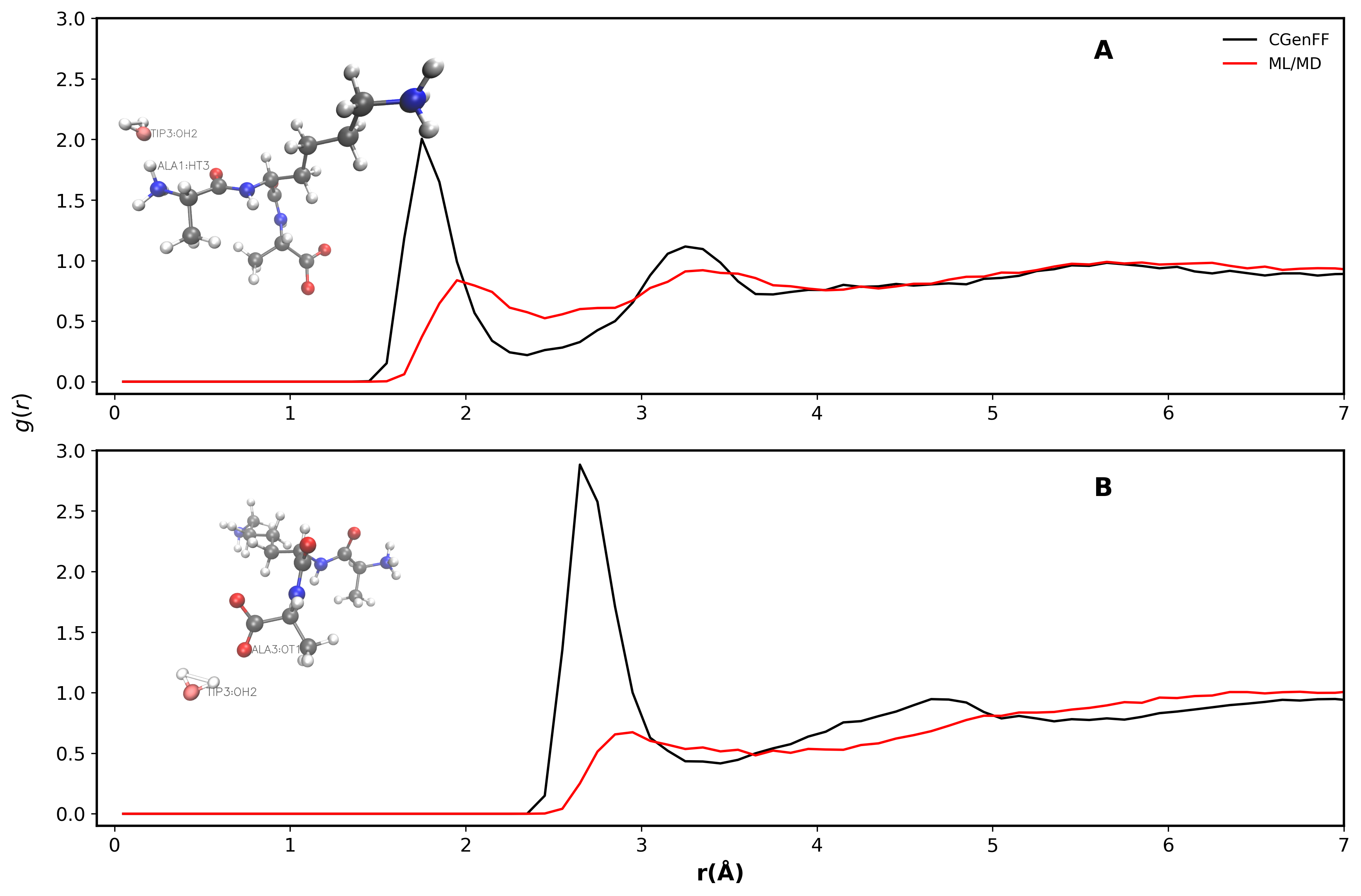}
\caption{\textbf{Radial distribution functions of the tripeptide AKA
    from CGenFF and ML-PES simulations.} $g(r)$ as a function of the
  distance $r$ between atom pairs O(H$_{2}$O)--H3(ALA1 residue
  -NH$_3^+$) (panel A) and O(H$_{2}$O)--O45(ALA3 residue -COO$^{-}$)
  (panel B) from simulations using CGenFF (black) and the ML-PES (red)
  simulation in solution.}
\label{fig:gr_sol}
\end{center}
\end{figure}

\noindent
Further analysis includes the calculation of the radial distribution
function (\textit{g(r)}) that describes how atomic or molecular
density varies relative to a reference particle. This analysis is
particularly useful for understanding the molecule-environment
interactions, and it is complementary to the previous analysis of IR
spectra. Figure \ref{fig:gr_sol}A shows $g(r)$ as a function of the
distance between O (from water molecules) and H3 (from the NH$_3^+$
group of the ALA1 residue) in the tripeptide. The simulations were
performed using CGenFF and the ML-PES (SETG5) model applied via
pyCHARMM. The first peak appears at 1.7 \AA\/ for the CGenFF
simulation and at 2.0 \AA\/ for the ML/MD simulation. The second shell
is observed around 3.2 \AA\/, though the peak in the ML/MD simulation
is slightly broader. The most significant difference is that the
CGenFF peaks have a much higher probability than the ML/MD simulation,
for which both peak maxima exceed 1.0. This confirms the previous
observations when comparing the charge distribution in Figure
\ref{fig:charges}. By examining, the partial charges of N1, H2, H3,
and H4, it is clear that the partial charge range of N1 in the ML/MD
simulation is higher than the CGenFF point charge, whereas the
hydrogen atoms exhibit slightly lower charges than in CGenFF. As a
result, when interacting with TIP3P water oxygen molecules, the
electrostatic attraction is slightly weaker in the ML/MD simulation,
also reflected in a broader second shell.  \\

\noindent
A similar observation can be made in Figure B, where $g(r)$ is
analyzed as a function of the distance between the atom pairs
water-oxygen atoms O$_{\rm W}$ and O45 (from the -COO$^{-}$ group of
the ALA3 residue). The first shell is observed around 2.6 and 2.9
\AA\/ for simulations using CGenFF and the ML-PES, respectively. The
second peak is broader in both simulations, but in the ML/MD
simulation, it shifts towards 5.0 \AA\/, whereas using CGenFF it
appears around 4.7 \AA\/. This decrease in peak intensity can be
explained by the less negative charge of the oxygen atom in the ML/MD
simulation compared to CGenFF (see Figure \ref{fig:charges}). This
decreases the electrostatic attraction between TIP3P water molecules
and the -COO$^{-}$ group. \\

\section{Application to a Reactive System: Proton Transfer in DNA Pairs}
\begin{figure}[h]
\centering
\includegraphics[width=1.0\textwidth]{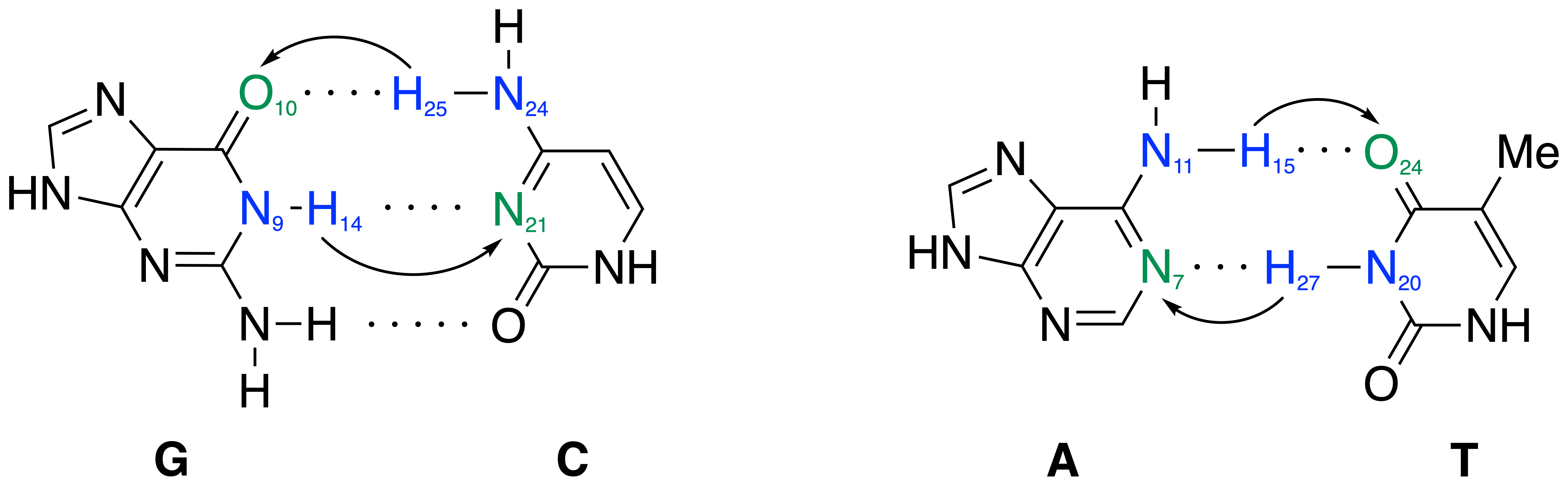}
\caption{\textbf{Double Proton Transfer scheme for GC (left) and AT
    (right)}.}
\label{fig:fig1}
\end{figure}

The second example to be discussed in more detail concerns the
construction of a reactive ML-PES. For this, double proton transfer in
DNA base pairs was considered. In 1963, L\"owdin proposed that
spontaneous mutations in DNA could occur as a result of double proton
transfer (DPT).\cite{lowdin:1963} In this process, see Figure
\ref{fig:fig1}, the Adenine-Thymine (AT) and Guanine-Cytosine(GC) base
pairs that make up DNA are transformed into their tautomeric forms:
A*T* and G*C*.\\
 
\noindent
One of the persistent open questions for doubly-H-bonded systems is
whether DPT is stepwise or synchronous,\cite{MM.da:2017,MM.da:2019}
and whether such a process can take place under
physiological/biological conditions with all base pairs in the
electronic ground state.\cite{slocombe:2022} Resolving these issues
poses significant challenges for computational chemistry because
standard density functional theory-based approaches are not
sufficiently accurate. Rather, high-level quantum chemical
calculations at the CCSD(T)/aug-cc-pVNZ (N = T or higher) are required
to obtain accurate H-transfer barrier heights.\cite{MM.tl:2025} One
possibility to reach such levels of theory is to generate a ML-PES at
a lower level of quantum chemical theory, e.g. density functional
theory or MP2, and then transfer learn it to a higher level of theory
using a limited subset of high-quality data.\cite{MM.tl:2022}\\

\noindent
Double Proton Transfer is an ideal process to illustrate how to
conceive and use ML-PESs for investigating fundamental and practically
important questions in chemistry and biology. The primary objective of
this example is to demonstrate, in a step-by-step manner, the process
of obtaining and validating an ML-PES for the base pairs. The model
may then be used in dynamic simulations to address specific
questions. First, the generation of initial geometries of all
reactants, products and monomers for both systems, AT and GC, in their
electronic ground state is described. This is followed by training and
validating the ML-PES and further improving it through transfer
learning.\\

\subsection{Exploration of the Systems}
Initially, optimised structures of the G, C, A, T monomers and the GC,
AT reactant and G*C*, A*T* product dimers were determined at the
B3LYP/def2-TZVP\cite{becke:1992,weigend:2005} level of theory together
with Grimme's dispersion correction D3BJ,\cite{grimme:2011} using the
ORCA 5.0.4 suite.\cite{ORCA5} For all systems, valid minimum energy
structures were found. However, although for the A*T* product state an
energy minimisation was possible, the nudged elastic band calculations
(see below) at this and other levels of theory indicated a monotonous
uphill energy profile without substantial stabilisation of the
product, which is consistent with previous findings\cite
{umesaki:2020}. Before constructing the ML-PES for the two dimers,
exploratory calculations were also carried out using the
computationally less demanding PBE/def2-SVP level of theory. Again,
for all species, minimum energy structures were found except for the
A*T* product state. At this lower level of theory, a local minimum for
A*T* was absent.\\

\begin{figure}[h!]
\centering
\includegraphics[width=0.75\textwidth]{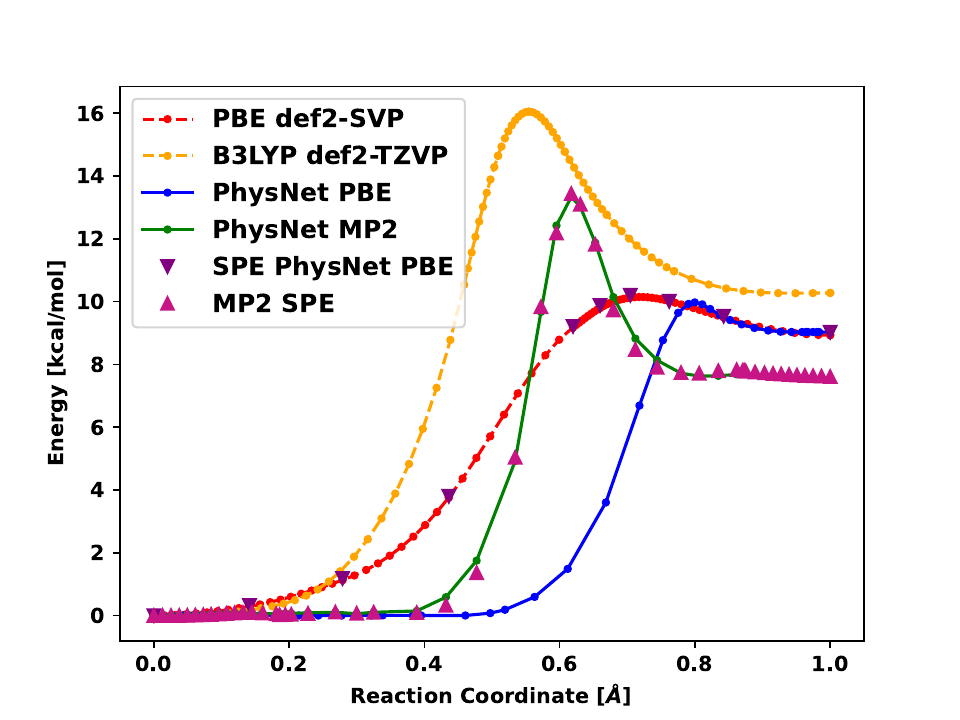}
\caption{\textbf{Nudged Elastic Band (NEB) profile for Double Proton
    Transfer in the pair GC.} The reaction profile was determined at
  the PBE/def2-SVP (red) and B3LYP/def2-TZVP (orange) level of theory
  and compared with the corresponding profiles obtained with the
  PhysNet model trained with data obtained at PBE/def2-SVP level of
  theory (blue). Complementary, the NEB path for the transfer learn to
  MP2/cc-pvTZ model is shown in green. Lastly, Single Point Energies
  were obtained for comparison in two cases. First, using the
  geometries obtained from the determined NEB path at PBE/def2-SVP
  evaluated with the obtained PhysNet model trained at the same level
  of theory (purple upside-down triangles). In the second case, a few
  geometries from the transfer learn model were evaluated at the
  MP2/cc-pvTZ (lila triangles). Notice that the reaction coordinates
  for the different profiles are individually scaled between 0 and 1
  and, therefore, not directly comparable. }
  \label{fig:fig2}
\end{figure}

\noindent
Complementary to the previous observations, changes in the equilibrium
geometry of the GC dimer were observed. At the B3LYP/def2-TZVP level
of theory, the GC dimer adopts a fully planar geometry. In contrast,
the optimised structure at the MP2/cc-pVTZ level displays a non-planar
geometry with the hydrogen atoms of the NH$_2$ group displaced out of
the molecular plane. Figure \ref{sifig:eq.geom} displays a comparison
of the optimised geometries of the GC dimer at the two levels of
theory, which feature a root-mean-square deviation (RMSD) of 0.084
\AA\/. Continuing with the characterisation of the system, Nudged
elastic band (NEB) calculations were carried out for the GC pair using
the two different density functionals (B3LYP and PBE) and the results
are reported in Figure \ref{fig:fig2}. The red and orange traces
correspond to the PBE/def2-SVP and B3LYP/def2-TZVP calculations,
respectively. The two methods found a stable geometry of the G*C*
product. However, the barrier heights and differential stabilization
of the products between the two levels of theory differ by 6 kcal/mol
and 1 kcal/mol, respectively. For AT, results from NEB calculations
are reported in Figure \ref{sifig:at_neb}. No stable A*T* geometry was
found at the B3LYP/def2-TZVP and MP2/cc-pVTZ levels of theory. \\

\subsection{Data Generation and Model Training}
To obtain an ML-PES to describe DPT for CG and AT, structures for
reference calculations were primarily generated through normal mode
sampling (NMS). This is different from the tripeptide example above,
for which molecular dynamics simulations were used. For the four
monomers (C, G, A, and T), structures were generated around the
PBE/def2-SVP minimum energy geometry using the normal mode vectors to
perform NMS. Conversely, for the four dimers (CG, C*G*, AT, A*T*), NEB
calculations using 8 images between the end points were performed at
the B3LYP/def2-TZVP level. For each of the obtained 10 structures, the
normal modes were determined at the PBE/def2-SVP level, and NMS was
carried out at 500 K. All structures were generated within the
Asparagus suite of codes\cite{MM.asparagus:2025} and using ORCA
5.0.4.\cite{ORCA5} This resulted in 9000 structures for each system,
CG and AT, including their monomers, which were used to train
PhysNet\cite{MM.physnet:2019} models using the Asparagus code.\\

\begin{figure}[h!]
\centering \includegraphics[width=1.0\textwidth]{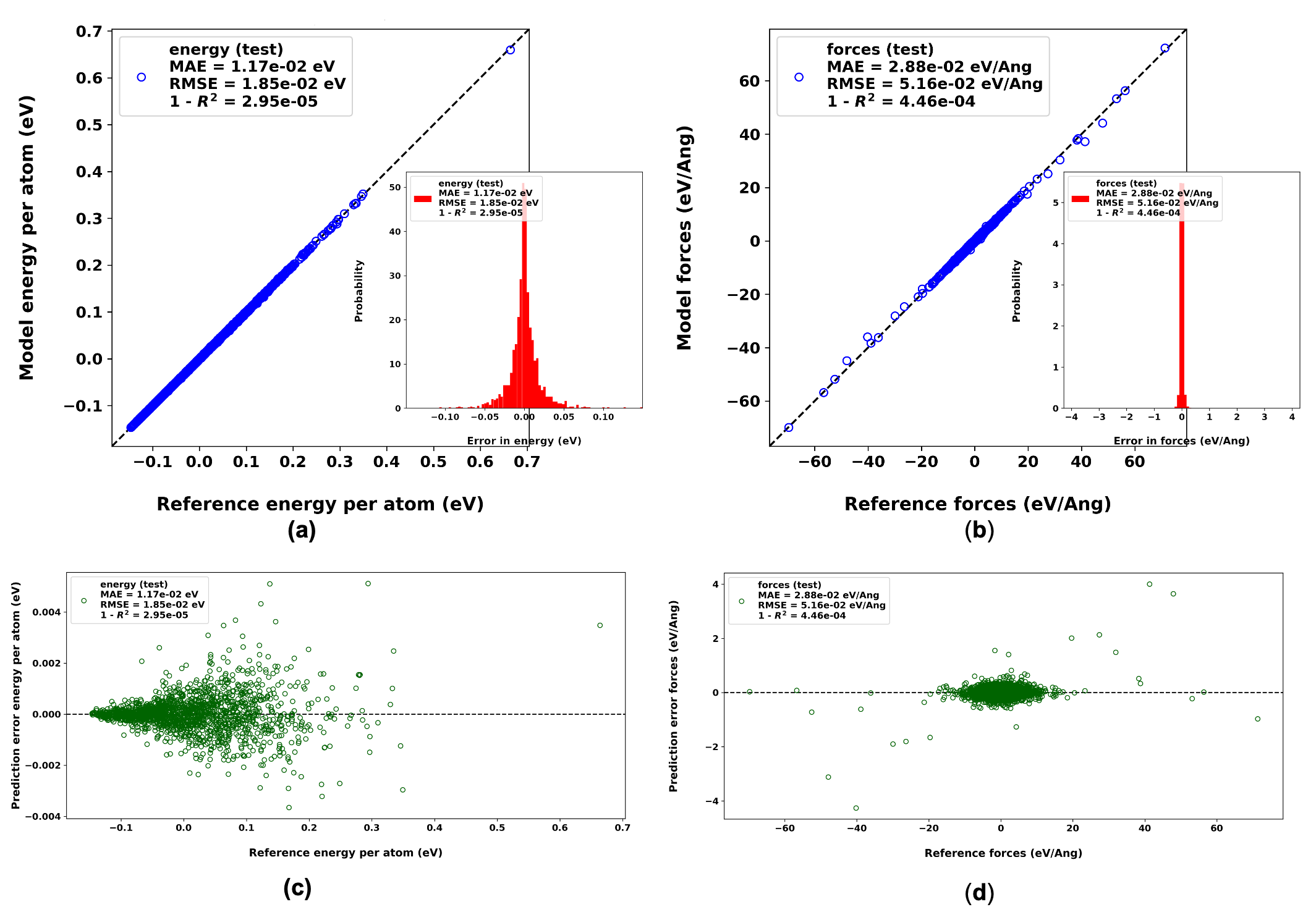}
\caption{\textbf{General performance of the ML model trained with
    PBE/def2-SVP} Panels A and B: Correlation of test set errors (main
  view) and error distributions (inset, red) for energies and forces
  for PhysNet Models trained on PBE/def2-SVP reference data. Panels C
  and D: Residual error for energies and forces for GC and AT
  basepairs: $x-$axis is reference $Value_{\rm PBE}$, $y-$axis is
  $\Delta Value = Value_{\rm PBE}-Value_{\rm PhysNet}$. The
  correlation between total molecular energy with reference to the
  global minimum is reported in Figure \ref{sifig:corr_plot_PBE}.}
\label{fig:fig3}
\end{figure}

\noindent
The final dataset contained 18000 reference energies, forces and
dipole moments, and was split into training, validation and test sets
with a ratio of 80:10:10, respectively. A first PhysNet model
comprising both systems, CG and AT, was then trained within the
Asparagus package.\cite{MM.physnet:2019,MM.asparagus:2025} The
correlation plots for energies and forces of the samples in the test
set, the corresponding error distributions, and residual plots are
illustrated in Figure \ref{fig:fig3}. The correlation coefficients
$(1-R^2)$ are $2.95 \cdot 10^{-5}$ for energies and $4.46 \cdot
10^{-4}$ for forces, with RMSE values of 0.43 kcal/mol for energies
and 1.19 (kcal/mol)$\cdot$ \AA$^{-1}$ for forces as well as MAE values
of 0.27 kcal/mol and 0.66 (kcal/mol)$\cdot$ \AA$^{-1}$ for forces,
which underline the high quality of the trained NNs. In addition, the
residual plots show most of the errors between -0.002 and 0.002 eV for
energies and -1 and 1 eV$\cdot$ \AA$^{-1}$ for forces. The previous
results demonstrate an excellent statistical performance in the test
set comparable with the results reported for the tripeptide system
discussed in the previous section, see Figure \ref{fig:fig3}. \\

\noindent
Next, the harmonic frequencies for the GC system were determined
directly from electronic structure calculations at the PBE/def2-SVP
level of theory. These are then compared with normal modes computed
from the trained ML-PES. For the optimised reactant, TS, and product
structures, the harmonic frequencies are summarised and compared
(column $\Delta$) in Table \ref{sitab:freq-pbe}. On average, the
frequencies differ by 7, 48, and 9 cm$^{-1}$ for the three geometries
with maximum deviations of 33, 1052, and 118 cm$^{-1}$ for the
reactant, TS, and product geometries, respectively. The differences
for the TS are, however, dominated by modes 72 and 73, which correspond
to vibrations involved in the proton transfer process. Comparison with the results
from the MP2 calculations (Table \ref{sitab:freq-tl}) indicates that
at the PBE level, this mode is not described sufficiently
well. Omitting these two modes when calculating the average deviations
between the ML-PES and the PBE reference method reduces the average
deviation to 22 cm$^{-1}$.\\

\noindent
For DPT, the relative stabilisation energies for reactants and
products and the barrier heights separating them are essential. For
both systems, results from the reference PBE/dev2-SVP calculations and
the trained PhysNet models were compared. To this end, optimizations
of reactants and products using the ASE\cite{larsen:2017} python
package were conducted with the Broyden–Fletcher–Goldfarb–Shanno
(BFGS) optimizer \cite{fletcher:2000}, with the maximum force
threshold value as termination criteria set to $5 \cdot 10^{-5}$
eV/\AA\/ allowed to perform NEB calculations. As illustrated in Figure
\ref{fig:fig2}, the NEB-profile (related to the minimum energy path) from
the ML-PES trained on PBE/dev2-SVP reference data (blue trace)
accurately predicts the barrier height and relative stabilization
computed at the PBE/dev2-SVP level (red trace). The deviation for the
transition state energy 10.12 kcal/mol, amounts to 0.35 kcal/mol. For
the AT system, both PBE/def2-SVP and the trained ML-PES do not find
a stable A*T* product.\\

\begin{figure}[htbp]
\centering
\includegraphics[width=1.0\textwidth]{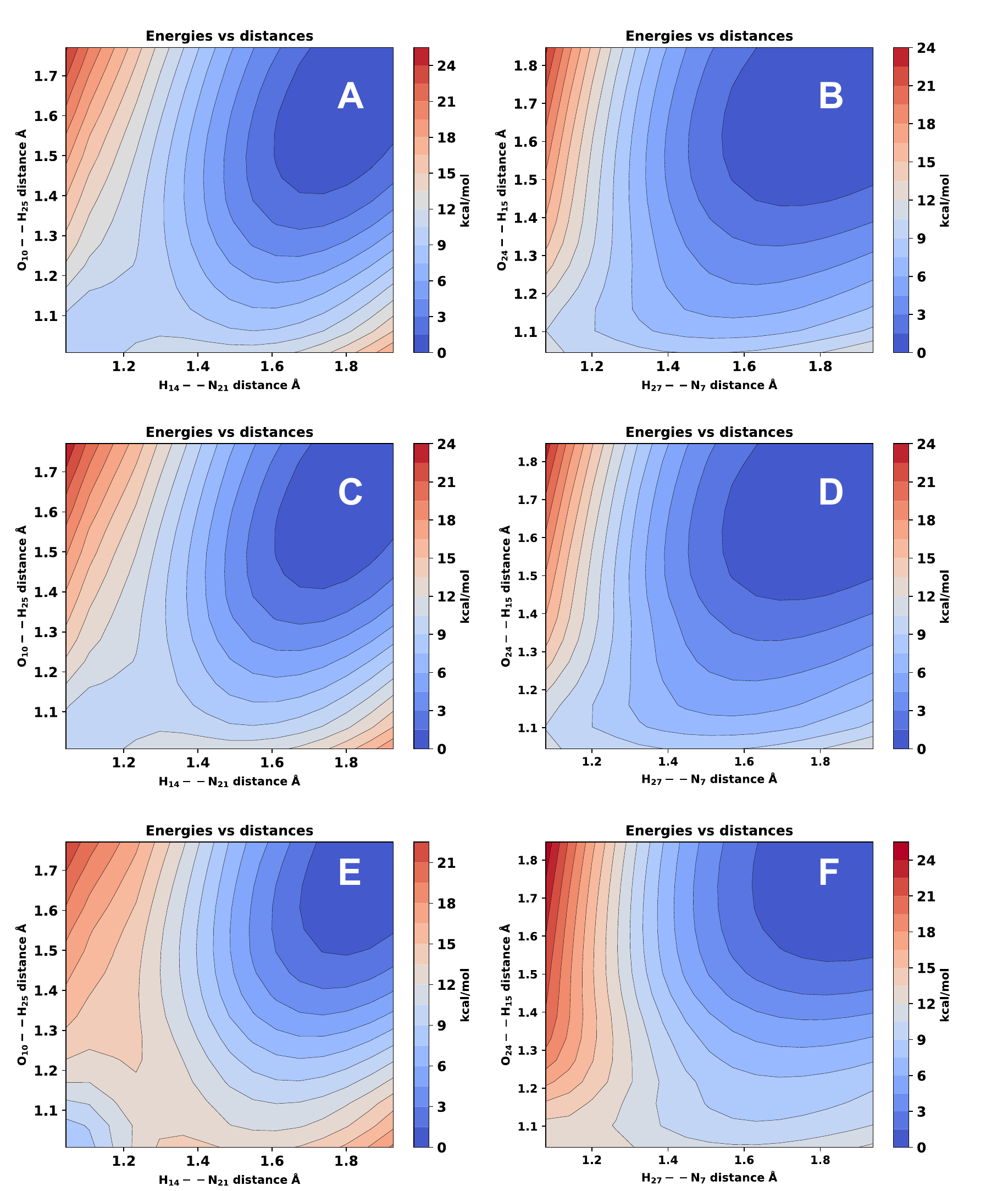}
\caption{\textbf{2D Potential Energy Surface for GC and AT base pairs}
  2D relaxed PES scans for the GC (left column) and AT (right column)
  basepairs were performed. The coordinates for scanning were the
  H14–-N21 and O10–-H25 separations for GC and the H27–-N7 and
  O24–-H15 separations for AT; for atom numbering, see Figure
  \ref{fig:fig1}. Panels A/B: The GC (A) and AT (B) pair at the
  PBE/def2-SVP level of theory. Panels C/D: The trained PhysNet models
  for GC (C) and AT (D). Panels E/F: The PhysNet models trained on
  MP2/cc-pvTZ level of theory for GC (E) and AT (F). Difference maps
  between reference energies and trained ML-PESs for both levels of
  theory are reported in Figure \ref{sifig:delta-maps}.}
\label{fig:fig5}
\end{figure}

\noindent
A yet more comprehensive comparison between reference data and trained
models consists of comparing relaxed 2-dimensional scans of the
PESs. For this, the H14–-N21 and O10–-H25 distances were used for the
GC pair. While for the AT pair, the coordinates considered were the
H27–-N7 and O24–-H15 distances. Figure \ref{fig:fig5}A/B compares the
relaxed 2d-PESs for GC, whereas \ref{fig:fig5}C/D are for the AT
pair. The correlation between model predictions and the reference
PBE/def2-SVP energies in the range evaluated are reported in Figures
\ref{sifig:corr_plot_PBE}A/B. The RMSE is 0.13 kcal/mol for both
systems and $1-R^2 = 0.0006$ for GC and 0.0007 for AT. Figures
\ref{fig:fig5} and \ref{sifig:corr_plot_PBE} underline the accuracy
and predictive power of the ML-PES, in particular as all these
validations were carried out for geometries that were not part of the
training set.\\

\noindent
As a final validation check of the ML-PESs, diffusion Monte-Carlo (DMC)
simulations as implemented in the Asparagus package were carried out
to identify ``holes''.\cite{qu:2021} For this purpose, 1000 walkers
were used for 1000 time steps, with a step size of $\Delta \tau = 5.0$
a.u. and a magnitude of fluctuations $\alpha = 0.1$. Evaluation of
close to $10^6$ structures along the DMC simulations did not find any
defective structure, which confirms the quality of the ML-PES and
suggests that stable and energy-conserving MD simulations can be run
in the future.\\

\subsection{Transfer Learning to MP2/cc-pVTZ}
As already mentioned, a realistic description of single- and double
proton transfer requires considerably higher levels of quantum
chemical theory than DFT-based methods. To this end, transfer learning
of the constructed ML-PES using DFT data to the MP2/cc-pVTZ level was
considered. For this procedure, a new, but considerably smaller
dataset was constructed based on a representative subset of the samples from the previous subsection. The dataset consisted of energies, forces, and dipole moments
for monomers (100 samples each, 400 in total), reactants and products
of the DPT process for AT and GC systems (100 samples each, 400 in
total), and 200 samples along each of the two NEB trajectories (400 in
total) calculated at the MP2/cc-pVTZ level of
theory.\cite{frisch:1990,kendall:1992} The TL-PES obtained by
retraining with 1,200 structures yields a barrier height for the
CG$\rightarrow$C*G* transition that differs by 0.63 kcal/mol from the
reference MP2/cc-pVTZ data.\\

\noindent
To further improve the model, 50 additional samples along each of the
two NEB trajectories (100 samples in total) were included. After
another round of transfer learning, the accuracy of the model further
improved, and the barrier heights difference to the reference values
differs by 0.52 kcal/mol. The final incorporation of an additional 500
samples for each NEB resulted in a total of 2300 additional data
points. These 1000 structures were generated from NEB trajectories
with perturbed structures by randomly displacing atoms with value of
standard deviation equal to 0.09 \AA\/ orthogonal to the NEB
pathway. Final TL reduced the difference between barrier heights for
DPT in GC at MP2/cc-pvTZ level of theory barrier height and PhysNet
model barrier height to 0.36 kcal/mol (12.82 kcal/mol for MP2/cc-pVTZ
and 13.18 kcal/mol for the trained PhysNet model) and the overall
performance is reported in Figure \ref{sifig:corr_plot_MP2}. To
complete the validation, normal modes for the GC system were
calculated using MP2/cc-pVTZ and the ML-PES for the reactant, TS and
product structures, see Table \ref{sitab:freq-tl}. On average, the
frequencies differ by 6, 9, and 8 cm$^{-1}$ for reactants, TS, and
product states. Most notably, the MAE for the TS is now comparable to
that of the reactant and product structures, which was not the case
when PBE/def2-SVP was used.\\

\noindent
It is noteworthy that the addition of samples along the CG and AT NEB
paths alone also enhanced the model's performance in predicting the
total energies of the monomers in their minimum energy
structures. Addition of 1200 and 2300 structures changes the energy
differences from --0.51 kcal/mol to +0.21 kcal/mol for adenine, --0.78
kcal/mol to +0.38 kcal/mol for guanine, --1.06 kcal/mol to +0.21
kcal/mol for cytosine, and --0.52 kcal/mol to +0.28 kcal/mol for
thymine. In other words, without explicitly including new data for the
monomers, the TL-PES improves the description of those species through
the addition of samples along the NEB paths.\\

\noindent
The constructed TL-PES was used to calculate the minimum dynamic
path\cite{unke2019sampling} (MDP) for the DPT in the CG pair (Figure
\ref{sifig:mdp_mep}). Starting from the TS, the molecule is relaxed
toward reactant or product. On the reactant part, it is observed that
large oscillations occur because of the formation of hydrogen bonds
with the carbonyl and amino groups, which produce collective
effects\cite{shi2017cooperative}. After $\sim$350 fs, the TS is
observed, followed by a very fast decay to the product (G*C*). Further
characterisation of the DPT transfer was obtained by studying the
distance between N and O atoms involved in the reaction as well as the
dihedral angles (Figure \ref{sifig:mdp_internal}). The interatomic
distances exhibit oscillatory behaviour, with a minimum of 2.5 \AA at
the TS. Small changes in the dihedral are observed, indicating that
the pair need to twist to favour the DPT. Finally, DMC simulations
along the same lines as for the ML-PES were carried out for CG and AT
using the TL-PES. Again, no defective structures were detected from
close to $10^6$ structures sampled during the DMC simulations, and the
TL-PESs are considered hole-free.\\

\noindent
The models discussed here can next be used in MD simulations to follow
DPT in the isolated dimers. Furthermore, these dimers can also be
embedded in larger oligomeric structures to study DPT in mixed ML/MM
simulations including hydration to account for environmental
effects.\cite{MM.DPT:2004}\\

\section{Open Questions, Challenges, and Perspectives}
The field of ML-PES is making progress very fast with new and
interesting application emerging. However, there are still some open
challenges that the field need to address. This section aims to
describe very briefly such aspects together with some of the current
challenges and present some perspectives. \\

\noindent
As discussed before, the construction of ML-PESs is an iterative
process where the model improves by adding new samples to the training
data. The addition is made base on uncertainty quantification
(UQ). This aspect considers and quantifies the expected accuracy of an
energy and force prediction when evaluating a trained ML-PES for
geometries that were not part of the training set. Several efforts
have been undertaken recently in this
direction.\cite{bombarelli:2023,MM.uq:2025} The finding was that UQ
using more or less sophisticated models for predicting uncertainty do,
in general, are inferior to predictions from ensembles of
models. Therefore, a clear challenge for ML-PES is the development of
quantitative models for UQ without unduly inflating training and
inference times while retaining accuracy and validity of the
uncertainty.\\

\noindent
Another open question concerns elevation of trained "base models" to
higher levels of quantum chemical theory. This step has been shown to
be required for quantitative atomistic simulations. Also, it should be
consider that f the available models are based on semi-empirical or
density functional theory data which in itself provide a great
accuracy but not enough for quantitative descriptions.  Currently, the
most used methods are are ``transfer
learning''\cite{MM.tl:2022,MM.tl:2025} and
``$\Delta-$learning''.\cite{lilienfeld:2015,nandi:2024} which have
been provided to be highly efficient. Furthermore with the appearance
of foundational models, the 'fine tuning' of such models to reach high
accuracies in specific tasks\cite{liu2025fine,radova2025fine} has
gained considerable interest. A key challenge for all of these
strategies is the selection of the structures at high level of theory
or that describe the desired task. Some progress indicate that
selecting points over the process of interest lead to accurate
description of it\cite{MM.tl:2022,deng2025systematic}. However,
further studies are required to find the best option to fulfill the
``dream'' of following the molecular dynamics of a protein in solution
at the CCSD(T) level of theory.\\

\noindent
It is also of much current and practical concern to minimize the
amount of reference data required for training reliable ML-PESs of
given quality for particular observables. Quantum chemistry
calculations at sufficiently high levels of theory are very time and
resource-intensive. Hence, keeping the reference data set small but
informative is essential. In this regard, an opening question is:
``Given $N$ reference energies/forces from reference calculations what
geometry should be used for the next calculation to maximally increase
the range of applicability of the resulting ML-PES.'' As an
alternative, this problem can be tackled with the use of generative
machine learning
models\cite{anstine2023generative,tiwary2025generative,dern2025energy}. \\

\noindent
Related to the previous point is the question of the most appropriate
reference system for which to calculate reference data. This is of
primary relevance for condensed phase simulations. Should reference
calculations be carried out separately for the solute and the solvent?
Or should a minimum number of solvent molecules always be present to
reflect electronic coupling between solute and solvent? Reference
calculation of the solute including a number of explicit solvent
molecules or in form of electrostatic embedded QM/MM simulations allow
the training of the highly accurate electrostatic embedded ML/MM PES
models with respect to external polarization of the ML system by the
MM environment. However, this increases the computational cost for the
reference data generation which may require to reduce the level of
quantum chemical theory.\\

\noindent
Nevertheless, the interaction potential also depends on the vdW
contribution often consisting of LJ potentials with pre-defined
parameters optimized in combination with simpler electrostatic models
such as fixed atom-centered charges for the ML atoms. When using the
fluctuating and externally polarized ML charges for the computation of
the ML-MM interaction potential, the LJ parameter needs to be adapted
to the new electrostatic model.  This means an LJ parameter
optimization with respect to reproduce best either experimental
condensed phase properties of the ML solute in the MM solvent or
reference interaction potential results from electronic structure
calculations. It is not necessarily possible to achieve good
agreements of multiple predicted experimental properties or reference
interaction potentials simultaneously.\\

\noindent
Finally, an important aspect of ML-PES is that the predictions
obtained fairly reproduce experimental observations. Therefore,
refinement of generated models with respect to experimental data is
still an open question. Efforts in this direction following simple
geometric transformation to 'morph' a ML-PES have been been
successfully applied to spectroscopic
applications\cite{MM.morphing:2024}. However, such methodology could
be expanded to consider other observables together with weighting
factors.\\

\section{Conclusions}
The successful and meaningful construction and
use of ML-PESs depend on a number of aspects that were explained and
illustrated in the text. To conclude, some general suggestions are
presented. First, when constructing and/or using a ML-PES, it is key to
have a clear definition of the purpose for which the ML-PES is
needed. Depending on this, the initial sampling, model selection, and further
refinements must be performed. As an example, a PES designed for
investigating chemical reactivity may not be equally suitable for
computing spectroscopic observables. This is in part related to the
fact that the two observables are sensitive to different areas of a
PES. It must be reemphasised that an ML-PES is a functional form that lacks a
physical basis; therefore, if the model enters into an 'extrapolation' regime, the obtained results would be wrong. On the other hand, the selection of the model should be based on the user's needs in terms of performance, accuracy, and computational budget.   \\

\noindent
The validation of a constructed ML-PES must not be limited to
statistical quantities such as RMSE or MAE. Further tests of
reliability must be performed to ensure the obtained model can be used
for running meaningful MD simulations. Some of them involve stability
measurements, calculation of radial distribution functions, using DMC
simulations, among others. It is also necessary to remember that
improvement in the prediction of reference values does not necessarily imply an improvement in the prediction of properties. \\

The combination of ML-PES with existing MD software opens the door for the simulation of larger and complex systems. Learning from the experiences of the QM/MM community is crucial for integrating and applying ML-PES in combination with established methods. As in the case of the model selection, the choice of an embedding scheme depends on the problem to be solved. The tripeptide example in Section 5 demonstrates how mixed methods yield significant improvements compared to classical force fields. \\

The field of ML-PES is advancing at a rapid yet stable pace, continually improving and expanding the data quantities, descriptions, and applications of models, with an increasing scope and impact. This work presented a 'hands-on' view of the field to motivate broader adoption of ML-based potential energy surfaces across disciplines. As the field advances, addressing challenges such as uncertainty quantification, data efficiency, integration of experimental observables, transferability, and model refinement will be crucial. Continued progress will depend on sustained collaboration and the shared pursuit of a better description of molecular systems. 

\section*{Research Data Availability Statement}
The data for the present study are available from
\url{https://github.com/MMunibas/practice} upon publication.

\section*{Acknowledgments}
The authors thank Dr. Silvan K\"aser for initial discussion about the
contents of this work. Financial support from the Swiss National
Science Foundation through grants $200020\_219779$ (MM),
$200021\_215088$ (MM), and the University of Basel (MM) is gratefully
acknowledged. LIVS acknowledges funding from the Swiss National
Science Foundation (Grant P500PN\_222297). KT acknowledges funding
by the Volkswagen Foundation through a Momentum grant.

\clearpage
\bibliography{refs}%
\clearpage
\appendix

\renewcommand{\thepage}{S\arabic{page}}
\renewcommand{\thetable}{S\arabic{table}}
\renewcommand{\thefigure}{S\arabic{figure}}
\renewcommand{\theequation}{S\arabic{equation}}

\setcounter{page}{1}
\setcounter{figure}{0}
\section{Tables}

\begin{longtable}{rrrrrrrrrr}
\caption{Harmonic frequencies (in cm$^{-1}$) for the CG pair, obtained from the Hessian matrix
($H=\partial^{2} E / \partial \boldsymbol{r}^{2}$) calculated using the PhysNet PES are compared
to their reference PBE/def2-SVP level of theory counterparts.}
\label{sitab:freq-pbe} \\
\toprule
 & \multicolumn{3}{c}{\textbf{Min}} & \multicolumn{3}{c}{\textbf{TS}} & \multicolumn{3}{c}{\textbf{Prod}} \\
\midrule
\textbf{  } & \textbf{PhysNet} & \textbf{PBE} & $|\Delta|$ & \textbf{PhysNet} & \textbf{PBE} & $|\Delta|$ & \textbf{PhysNet} & \textbf{PBE} & $|\Delta|$ \\
\midrule
\endfirsthead
\toprule
\textbf{  } & \textbf{PhysNet} & \textbf{PBE} & $|\Delta|$ & \textbf{PhysNet} & \textbf{PBE} & $|\Delta|$ & \textbf{PhysNet} & \textbf{PBE} & $|\Delta|$ \\
\midrule
\endhead
\midrule
\\
\endfoot
\bottomrule
\endlastfoot

1	&	24.80	&	27.71	&	2.91	&	-1009.63	&	-346.35	&		&	19.10	&	13.29	&	5.81	\\
2	&	42.80	&	35.82	&	6.98	&	20.81	&	25.13	&	4.32	&	33.10	&	31.04	&	2.06	\\
3	&	78.50	&	77.97	&	0.53	&	30.65	&	40.41	&	9.76	&	77.40	&	82.01	&	4.61	\\
4	&	102.90	&	102.93	&	0.03	&	78.90	&	70.69	&	8.21	&	124.50	&	121.82	&	2.68	\\
5	&	137.30	&	133.17	&	4.13	&	125.14	&	95.83	&	29.31	&	131.80	&	123.17	&	8.63	\\
6	&	143.20	&	143.98	&	0.78	&	131.74	&	122.50	&	9.24	&	137.00	&	135.89	&	1.11	\\
7	&	153.10	&	149.57	&	3.53	&	158.77	&	141.79	&	16.98	&	160.70	&	159.19	&	1.51	\\
8	&	166.20	&	155.53	&	10.67	&	168.62	&	146.29	&	22.33	&	167.60	&	162.37	&	5.23	\\
9	&	178.90	&	178.03	&	0.87	&	183.82	&	160.13	&	23.69	&	175.80	&	169.23	&	6.57	\\
10	&	198.50	&	193.51	&	4.99	&	191.30	&	178.77	&	12.53	&	187.30	&	193.66	&	6.36	\\
11	&	215.60	&	216.11	&	0.51	&	252.11	&	198.11	&	54.00	&	250.60	&	246.61	&	3.99	\\
12	&	331.50	&	342.56	&	11.06	&	253.23	&	229.95	&	23.28	&	323.50	&	335.73	&	12.23	\\
13	&	348.40	&	347.45	&	0.95	&	326.72	&	338.31	&	11.59	&	347.60	&	343.35	&	4.25	\\
14	&	380.30	&	379.51	&	0.79	&	356.45	&	361.23	&	4.78	&	373.60	&	395.18	&	21.58	\\
15	&	400.60	&	399.87	&	0.73	&	401.51	&	371.12	&	30.39	&	393.80	&	400.12	&	6.32	\\
16	&	407.60	&	418.77	&	11.17	&	402.15	&	399.36	&	2.79	&	398.20	&	404.97	&	6.77	\\
17	&	420.50	&	419.49	&	1.01	&	416.67	&	414.10	&	2.57	&	440.60	&	433.76	&	6.84	\\
18	&	491.20	&	491.34	&	0.14	&	462.78	&	487.01	&	24.23	&	499.30	&	496.47	&	2.83	\\
19	&	494.60	&	496.64	&	2.04	&	499.57	&	487.85	&	11.72	&	506.90	&	533.15	&	26.25	\\
20	&	526.50	&	527.54	&	1.04	&	512.18	&	531.76	&	19.58	&	533.80	&	545.63	&	11.83	\\
21	&	532.20	&	539.74	&	7.54	&	541.98	&	535.50	&	6.48	&	548.10	&	546.24	&	1.86	\\
22	&	539.30	&	556.18	&	16.88	&	554.53	&	550.42	&	4.11	&	557.00	&	559.33	&	2.33	\\
23	&	558.90	&	556.77	&	2.13	&	573.58	&	575.69	&	2.11	&	569.70	&	562.89	&	6.81	\\
24	&	579.90	&	579.26	&	0.64	&	582.33	&	581.31	&	1.02	&	570.70	&	572.80	&	2.10	\\
25	&	583.30	&	616.70	&	33.40	&	607.15	&	621.23	&	14.08	&	620.50	&	634.11	&	13.61	\\
26	&	627.30	&	637.94	&	10.64	&	623.22	&	626.42	&	3.20	&	634.80	&	643.60	&	8.80	\\
27	&	639.70	&	647.00	&	7.30	&	635.82	&	644.03	&	8.21	&	646.10	&	647.96	&	1.86	\\
28	&	654.70	&	665.02	&	10.32	&	642.89	&	662.36	&	19.47	&	650.40	&	679.79	&	29.39	\\
29	&	669.40	&	680.65	&	11.25	&	658.05	&	686.52	&	28.47	&	668.00	&	686.06	&	18.06	\\
30	&	684.00	&	701.92	&	17.92	&	671.55	&	694.23	&	22.68	&	671.70	&	695.90	&	24.20	\\
31	&	692.80	&	706.25	&	13.45	&	698.63	&	703.90	&	5.27	&	698.80	&	712.21	&	13.41	\\
32	&	732.10	&	737.33	&	5.23	&	710.27	&	751.71	&	41.44	&	728.50	&	733.00	&	4.50	\\
33	&	742.90	&	753.51	&	10.61	&	734.51	&	758.47	&	23.96	&	733.90	&	744.77	&	10.87	\\
34	&	755.50	&	757.44	&	1.94	&	740.18	&	768.69	&	28.51	&	756.90	&	767.47	&	10.57	\\
35	&	761.30	&	759.84	&	1.46	&	759.86	&	774.29	&	14.43	&	768.90	&	770.09	&	1.19	\\
36	&	774.20	&	773.84	&	0.36	&	768.54	&	802.73	&	34.19	&	772.10	&	773.38	&	1.28	\\
37	&	783.40	&	783.88	&	0.48	&	792.12	&	821.31	&	29.19	&	792.70	&	792.49	&	0.21	\\
38	&	821.60	&	819.93	&	1.67	&	794.16	&	906.59	&	112.43	&	831.10	&	827.18	&	3.92	\\
39	&	898.20	&	919.29	&	21.09	&	839.89	&	911.93	&	72.04	&	896.10	&	919.12	&	23.02	\\
40	&	921.50	&	924.69	&	3.19	&	897.69	&	922.86	&	25.17	&	921.10	&	922.75	&	1.65	\\
41	&	923.30	&	936.32	&	13.02	&	922.75	&	925.76	&	3.01	&	953.40	&	954.28	&	0.88	\\
42	&	947.60	&	952.11	&	4.51	&	993.84	&	945.45	&	48.39	&	1000.10	&	996.82	&	3.28	\\
43	&	952.40	&	972.87	&	20.47	&	1003.40	&	981.85	&	21.55	&	1039.10	&	1036.67	&	2.43	\\
44	&	984.70	&	983.51	&	1.19	&	1051.73	&	1010.72	&	41.01	&	1043.10	&	1044.37	&	1.27	\\
45	&	1046.90	&	1038.29	&	8.61	&	1065.15	&	1044.22	&	20.93	&	1053.90	&	1051.11	&	2.79	\\
46	&	1054.30	&	1041.68	&	12.62	&	1078.03	&	1058.41	&	19.62	&	1073.20	&	1073.29	&	0.09	\\
47	&	1077.60	&	1085.28	&	7.68	&	1090.95	&	1088.11	&	2.84	&	1079.00	&	1080.76	&	1.76	\\
48	&	1089.50	&	1088.08	&	1.42	&	1124.41	&	1100.06	&	24.35	&	1102.90	&	1110.81	&	7.91	\\
49	&	1104.00	&	1103.46	&	0.54	&	1139.69	&	1132.02	&	7.67	&	1138.10	&	1139.13	&	1.03	\\
50	&	1155.00	&	1147.64	&	7.36	&	1158.92	&	1169.63	&	10.71	&	1152.90	&	1147.20	&	5.70	\\
51	&	1184.00	&	1180.89	&	3.11	&	1194.72	&	1181.69	&	13.03	&	1195.30	&	1184.55	&	10.75	\\
52	&	1189.70	&	1184.60	&	5.10	&	1205.75	&	1185.19	&	20.56	&	1227.80	&	1232.68	&	4.88	\\
53	&	1287.80	&	1290.40	&	2.60	&	1223.55	&	1279.74	&	56.19	&	1299.30	&	1309.46	&	10.16	\\
54	&	1297.60	&	1303.31	&	5.71	&	1301.38	&	1283.04	&	18.34	&	1305.60	&	1311.67	&	6.07	\\
55	&	1331.40	&	1332.78	&	1.38	&	1304.61	&	1289.10	&	15.51	&	1348.00	&	1344.51	&	3.49	\\
56	&	1349.50	&	1346.95	&	2.55	&	1339.95	&	1330.49	&	9.46	&	1350.80	&	1351.44	&	0.64	\\
57	&	1363.30	&	1364.50	&	1.20	&	1348.10	&	1354.22	&	6.12	&	1360.60	&	1359.18	&	1.42	\\
58	&	1393.60	&	1396.46	&	2.86	&	1354.29	&	1374.72	&	20.43	&	1389.70	&	1386.99	&	2.71	\\
59	&	1414.60	&	1405.75	&	8.85	&	1372.90	&	1408.87	&	35.97	&	1410.20	&	1411.77	&	1.57	\\
60	&	1445.10	&	1438.07	&	7.03	&	1402.36	&	1428.88	&	26.52	&	1428.10	&	1438.28	&	10.18	\\
61	&	1491.80	&	1502.43	&	10.63	&	1430.44	&	1485.70	&	55.26	&	1457.80	&	1464.96	&	7.16	\\
62	&	1512.00	&	1507.42	&	4.58	&	1477.30	&	1504.75	&	27.45	&	1493.10	&	1491.82	&	1.28	\\
63	&	1530.10	&	1530.17	&	0.07	&	1504.68	&	1518.38	&	13.70	&	1500.50	&	1499.21	&	1.29	\\
64	&	1540.80	&	1535.11	&	5.69	&	1516.53	&	1529.95	&	13.42	&	1504.70	&	1511.47	&	6.77	\\
65	&	1579.80	&	1583.10	&	3.30	&	1524.36	&	1557.38	&	33.02	&	1586.30	&	1566.05	&	20.25	\\
66	&	1605.50	&	1609.45	&	3.95	&	1590.62	&	1578.64	&	11.98	&	1596.50	&	1609.75	&	13.25	\\
67	&	1631.40	&	1619.60	&	11.80	&	1611.55	&	1616.86	&	5.31	&	1607.60	&	1622.42	&	14.82	\\
68	&	1650.00	&	1656.20	&	6.20	&	1622.73	&	1622.48	&	0.25	&	1642.60	&	1639.46	&	3.14	\\
69	&	1684.40	&	1679.33	&	5.07	&	1649.05	&	1674.49	&	25.44	&	1654.00	&	1657.26	&	3.26	\\
70	&	1715.20	&	1714.07	&	1.13	&	1667.54	&	1723.61	&	56.07	&	1702.40	&	1702.17	&	0.23	\\
71	&	1758.60	&	1747.26	&	11.34	&	1716.73	&	1757.77	&	41.04	&	1774.50	&	1770.38	&	4.12	\\
72	&	2700.90	&	2723.77	&	22.87	&	1774.68	&	2788.06	&	1013.38	&	2088.60	&	2206.87	&	118.27	\\
73	&	2924.90	&	2912.75	&	12.15	&	1832.50	&	2884.75	&	1052.25	&	2382.60	&	2425.41	&	42.81	\\
74	&	3135.60	&	3140.06	&	4.46	&	3137.53	&	3137.17	&	0.36	&	3131.50	&	3144.07	&	12.57	\\
75	&	3152.50	&	3163.30	&	10.80	&	3159.29	&	3165.23	&	5.94	&	3155.80	&	3161.15	&	5.35	\\
76	&	3161.00	&	3165.44	&	4.44	&	3177.16	&	3165.48	&	11.68	&	3181.20	&	3173.12	&	8.08	\\
77	&	3173.40	&	3194.59	&	21.19	&	3218.80	&	3347.68	&	128.88	&	3322.60	&	3325.67	&	3.07	\\
78	&	3530.30	&	3538.77	&	8.47	&	3485.67	&	3439.07	&	46.60	&	3467.30	&	3476.78	&	9.48	\\
79	&	3540.60	&	3560.94	&	20.34	&	3537.11	&	3537.07	&	0.04	&	3541.30	&	3555.48	&	14.18	\\
80	&	3574.30	&	3586.80	&	12.50	&	3546.09	&	3554.68	&	8.59	&	3546.90	&	3564.74	&	17.84	\\
81	&	3583.30	&	3611.64	&	28.34	&	3577.50	&	3589.66	&	12.16	&	3595.80	&	3621.21	&	25.41	\\\midrule
\textbf{MAE}&  \textbf{7.03} & & & \textbf{47.53} & & & \textbf{9.12} &&\\\bottomrule

\end{longtable}

\clearpage

\begin{longtable}{rrrrrrrrrr}
\caption{Harmonic frequencies for the CG pair, obtained from the Hessian matrix
($H=\partial^{2} E / \partial \boldsymbol{r}^{2}$) calculated using the PhysNet PES are compared
to their \textit{ab initio} MP2/cc-pVTZ level of theory counterparts.}
\label{sitab:freq-tl} \\
\toprule
 & \multicolumn{3}{c}{\textbf{Min}} & \multicolumn{3}{c}{\textbf{TS}} & \multicolumn{3}{c}{\textbf{Prod}} \\
\midrule
\textbf{  } & \textbf{PhysNet} & \textbf{MP2} & $|\Delta|$ & \textbf{PhysNet} & \textbf{MP2} & $|\Delta|$ & \textbf{PhysNet} & \textbf{MP2} & $|\Delta|$ \\
\midrule
\endfirsthead
\toprule
\textbf{  } & \textbf{PhysNet} & \textbf{MP2} & $|\Delta|$ & \textbf{PhysNet} & \textbf{MP2} & $|\Delta|$ & \textbf{PhysNet} & \textbf{MP2} & $|\Delta|$ \\
\midrule
\endhead
\midrule
\\
\endfoot
\bottomrule
\endlastfoot

1	&	25.29	&	25.16	&	0.13	&	-1225.00	&	-1210.12	&		&	28.04	&	29.52	&	1.48	\\
2	&	40.70	&	35.81	&	4.89	&	27.16	&	34.09	&	6.93	&	35.61	&	36.06	&	0.45	\\
3	&	74.05	&	69.15	&	4.90	&	39.81	&	47.03	&	7.22	&	72.35	&	75.61	&	3.26	\\
4	&	93.99	&	89.30	&	4.69	&	76.75	&	79.90	&	3.15	&	103.80	&	110.35	&	6.55	\\
5	&	121.39	&	130.68	&	9.29	&	119.81	&	118.11	&	1.70	&	116.08	&	117.73	&	1.65	\\
6	&	127.09	&	137.05	&	9.96	&	146.32	&	141.32	&	5.00	&	130.89	&	139.80	&	8.91	\\
7	&	149.54	&	145.34	&	4.20	&	161.34	&	151.14	&	10.20	&	152.07	&	142.36	&	9.71	\\
8	&	170.18	&	156.79	&	13.39	&	174.90	&	157.38	&	17.52	&	153.29	&	153.36	&	0.07	\\
9	&	184.05	&	178.84	&	5.21	&	189.12	&	187.69	&	1.43	&	168.77	&	164.05	&	4.72	\\
10	&	198.56	&	199.11	&	0.55	&	196.90	&	198.49	&	1.59	&	197.39	&	200.92	&	3.53	\\
11	&	219.86	&	217.94	&	1.92	&	244.67	&	248.50	&	3.83	&	256.35	&	253.38	&	2.97	\\
12	&	339.05	&	343.76	&	4.71	&	262.86	&	258.14	&	4.72	&	333.15	&	339.61	&	6.46	\\
13	&	346.52	&	358.58	&	12.06	&	345.18	&	330.07	&	15.11	&	348.77	&	355.99	&	7.22	\\
14	&	369.60	&	371.50	&	1.90	&	348.40	&	356.99	&	8.59	&	385.99	&	389.94	&	3.95	\\
15	&	414.28	&	400.56	&	13.72	&	409.60	&	398.94	&	10.66	&	405.71	&	392.89	&	12.82	\\
16	&	417.79	&	410.37	&	7.42	&	411.26	&	404.99	&	6.27	&	417.22	&	415.84	&	1.38	\\
17	&	435.15	&	443.50	&	8.35	&	469.56	&	456.01	&	13.55	&	452.40	&	500.29	&	47.89	\\
18	&	456.82	&	446.12	&	10.70	&	488.90	&	500.26	&	11.36	&	488.64	&	507.13	&	18.49	\\
19	&	497.81	&	499.52	&	1.71	&	500.50	&	526.79	&	26.29	&	507.62	&	529.21	&	21.59	\\
20	&	521.74	&	536.28	&	14.54	&	509.19	&	537.58	&	28.39	&	538.00	&	534.23	&	3.77	\\
21	&	536.00	&	546.92	&	10.92	&	549.95	&	543.57	&	6.38	&	554.94	&	550.19	&	4.75	\\
22	&	544.73	&	547.14	&	2.41	&	561.20	&	558.93	&	2.27	&	556.57	&	555.25	&	1.32	\\
23	&	562.21	&	558.31	&	3.90	&	572.18	&	574.84	&	2.66	&	568.06	&	557.76	&	10.30	\\
24	&	584.32	&	586.51	&	2.19	&	586.61	&	583.81	&	2.80	&	570.39	&	569.71	&	0.68	\\
25	&	594.62	&	623.80	&	29.18	&	610.49	&	612.34	&	1.85	&	592.82	&	603.39	&	10.57	\\
26	&	634.20	&	628.16	&	6.04	&	642.67	&	634.85	&	7.82	&	638.63	&	642.04	&	3.41	\\
27	&	645.09	&	644.09	&	1.00	&	647.42	&	662.48	&	15.06	&	639.31	&	661.97	&	22.66	\\
28	&	645.55	&	666.37	&	20.82	&	667.84	&	682.92	&	15.08	&	680.85	&	693.13	&	12.28	\\
29	&	684.34	&	684.69	&	0.35	&	687.49	&	693.78	&	6.29	&	686.35	&	696.98	&	10.63	\\
30	&	689.42	&	701.59	&	12.17	&	696.90	&	706.10	&	9.20	&	698.22	&	706.51	&	8.29	\\
31	&	700.04	&	730.38	&	30.34	&	712.82	&	724.88	&	12.06	&	704.18	&	741.57	&	37.39	\\
32	&	727.57	&	739.53	&	11.96	&	732.84	&	733.51	&	0.67	&	738.79	&	753.89	&	15.10	\\
33	&	775.43	&	780.91	&	5.48	&	737.91	&	751.05	&	13.14	&	751.34	&	769.23	&	17.89	\\
34	&	783.65	&	781.05	&	2.60	&	767.95	&	776.58	&	8.63	&	783.17	&	785.41	&	2.24	\\
35	&	789.09	&	785.73	&	3.36	&	787.38	&	788.45	&	1.07	&	785.64	&	800.67	&	15.03	\\
36	&	789.85	&	790.38	&	0.53	&	793.72	&	799.49	&	5.77	&	795.36	&	803.31	&	7.95	\\
37	&	806.72	&	820.01	&	13.29	&	804.99	&	808.18	&	3.19	&	817.76	&	836.47	&	18.71	\\
38	&	831.97	&	834.39	&	2.42	&	819.35	&	835.89	&	16.54	&	838.01	&	840.51	&	2.50	\\
39	&	845.50	&	855.97	&	10.47	&	857.88	&	858.76	&	0.88	&	927.85	&	936.05	&	8.20	\\
40	&	899.15	&	911.22	&	12.07	&	919.82	&	938.29	&	18.47	&	939.06	&	943.48	&	4.42	\\
41	&	934.39	&	937.57	&	3.18	&	941.98	&	956.12	&	14.14	&	941.89	&	955.04	&	13.15	\\
42	&	942.49	&	960.92	&	18.43	&	1010.14	&	996.50	&	13.64	&	974.33	&	981.55	&	7.22	\\
43	&	971.22	&	974.99	&	3.77	&	1032.47	&	1027.87	&	4.60	&	989.14	&	991.32	&	2.18	\\
44	&	1008.68	&	1006.28	&	2.40	&	1079.90	&	1078.00	&	1.90	&	1017.93	&	1016.11	&	1.82	\\
45	&	1072.63	&	1068.14	&	4.49	&	1100.61	&	1087.02	&	13.59	&	1054.43	&	1055.70	&	1.27	\\
46	&	1099.95	&	1089.77	&	10.18	&	1118.74	&	1105.71	&	13.03	&	1093.65	&	1091.14	&	2.51	\\
47	&	1124.18	&	1123.65	&	0.53	&	1143.41	&	1121.14	&	22.27	&	1108.32	&	1107.98	&	0.34	\\
48	&	1132.86	&	1128.35	&	4.51	&	1149.98	&	1147.89	&	2.09	&	1123.85	&	1123.98	&	0.13	\\
49	&	1143.15	&	1146.77	&	3.62	&	1190.76	&	1187.37	&	3.39	&	1183.08	&	1175.04	&	8.04	\\
50	&	1187.41	&	1177.58	&	9.83	&	1197.90	&	1202.81	&	4.91	&	1189.05	&	1181.46	&	7.59	\\
51	&	1218.31	&	1214.89	&	3.42	&	1234.22	&	1231.01	&	3.21	&	1232.26	&	1224.65	&	7.61	\\
52	&	1240.12	&	1226.93	&	13.19	&	1246.31	&	1250.72	&	4.41	&	1251.64	&	1251.96	&	0.32	\\
53	&	1311.12	&	1312.52	&	1.40	&	1258.12	&	1260.76	&	2.64	&	1323.01	&	1324.88	&	1.87	\\
54	&	1324.02	&	1332.34	&	8.32	&	1330.52	&	1310.61	&	19.91	&	1347.96	&	1353.51	&	5.55	\\
55	&	1384.38	&	1389.18	&	4.80	&	1340.80	&	1330.98	&	9.82	&	1385.90	&	1393.18	&	7.28	\\
56	&	1388.87	&	1391.85	&	2.98	&	1371.99	&	1358.83	&	13.16	&	1394.19	&	1401.94	&	7.75	\\
57	&	1402.47	&	1404.53	&	2.06	&	1390.37	&	1396.14	&	5.77	&	1405.45	&	1411.71	&	6.26	\\
58	&	1421.34	&	1424.53	&	3.19	&	1405.44	&	1408.34	&	2.90	&	1413.62	&	1418.69	&	5.07	\\
59	&	1453.84	&	1452.28	&	1.56	&	1406.12	&	1413.33	&	7.21	&	1463.14	&	1458.72	&	4.42	\\
60	&	1460.72	&	1466.81	&	6.09	&	1434.80	&	1441.34	&	6.54	&	1465.45	&	1477.73	&	12.28	\\
61	&	1510.19	&	1509.75	&	0.44	&	1462.36	&	1460.88	&	1.48	&	1465.89	&	1480.43	&	14.54	\\
62	&	1542.18	&	1554.07	&	11.89	&	1476.83	&	1483.03	&	6.20	&	1509.40	&	1517.59	&	8.19	\\
63	&	1565.47	&	1572.55	&	7.08	&	1538.32	&	1545.97	&	7.65	&	1526.98	&	1529.77	&	2.79	\\
64	&	1578.52	&	1583.48	&	4.96	&	1544.29	&	1555.64	&	11.35	&	1533.74	&	1533.96	&	0.22	\\
65	&	1613.77	&	1624.89	&	11.12	&	1556.23	&	1564.94	&	8.71	&	1565.91	&	1560.87	&	5.04	\\
66	&	1657.79	&	1663.64	&	5.85	&	1620.76	&	1623.49	&	2.73	&	1639.96	&	1652.85	&	12.89	\\
67	&	1678.51	&	1679.70	&	1.19	&	1633.00	&	1645.82	&	12.82	&	1654.58	&	1666.03	&	11.45	\\
68	&	1687.98	&	1695.29	&	7.31	&	1674.38	&	1684.98	&	10.60	&	1670.49	&	1676.82	&	6.33	\\
69	&	1730.72	&	1725.92	&	4.80	&	1702.82	&	1693.50	&	9.32	&	1694.33	&	1698.75	&	4.42	\\
70	&	1744.71	&	1747.70	&	2.99	&	1717.01	&	1723.21	&	6.20	&	1735.36	&	1736.98	&	1.62	\\
71	&	1777.70	&	1779.46	&	1.76	&	1743.90	&	1736.61	&	7.29	&	1802.49	&	1804.61	&	2.12	\\
72	&	3117.79	&	3121.08	&	3.29	&	1813.26	&	1749.63	&	63.63	&	2933.45	&	2909.61	&	23.84	\\
73	&	3191.77	&	3187.57	&	4.20	&	1853.10	&	1821.27	&	31.83	&	3031.35	&	3021.51	&	9.84	\\
74	&	3246.86	&	3244.54	&	2.32	&	3241.81	&	3242.15	&	0.34	&	3242.92	&	3245.02	&	2.10	\\
75	&	3272.48	&	3271.26	&	1.22	&	3265.67	&	3269.59	&	3.92	&	3262.52	&	3273.80	&	11.28	\\
76	&	3283.50	&	3284.17	&	0.67	&	3294.66	&	3282.86	&	11.80	&	3290.40	&	3281.02	&	9.38	\\
77	&	3417.85	&	3410.87	&	6.98	&	3363.17	&	3391.17	&	28.00	&	3517.56	&	3506.80	&	10.76	\\
78	&	3646.96	&	3647.80	&	0.84	&	3593.53	&	3595.61	&	2.08	&	3576.15	&	3556.20	&	19.95	\\
79	&	3674.54	&	3673.56	&	0.98	&	3659.76	&	3661.00	&	1.24	&	3675.44	&	3668.54	&	6.90	\\
80	&	3709.77	&	3710.99	&	1.22	&	3670.09	&	3673.98	&	3.89	&	3676.83	&	3675.86	&	0.97	\\
81	&	3715.03	&	3714.51	&	0.52	&	3675.23	&	3686.44	&	11.21	&	3705.78	&	3701.50	&	4.28	\\
\\\midrule
\textbf{MAE}&  \textbf{6.24} & & & \textbf{9.31} & & & \textbf{8.06} &&\\\bottomrule

\end{longtable}

\clearpage
\section{Figures}
\begin{figure}[htbp]
\includegraphics[width=0.75\textwidth]{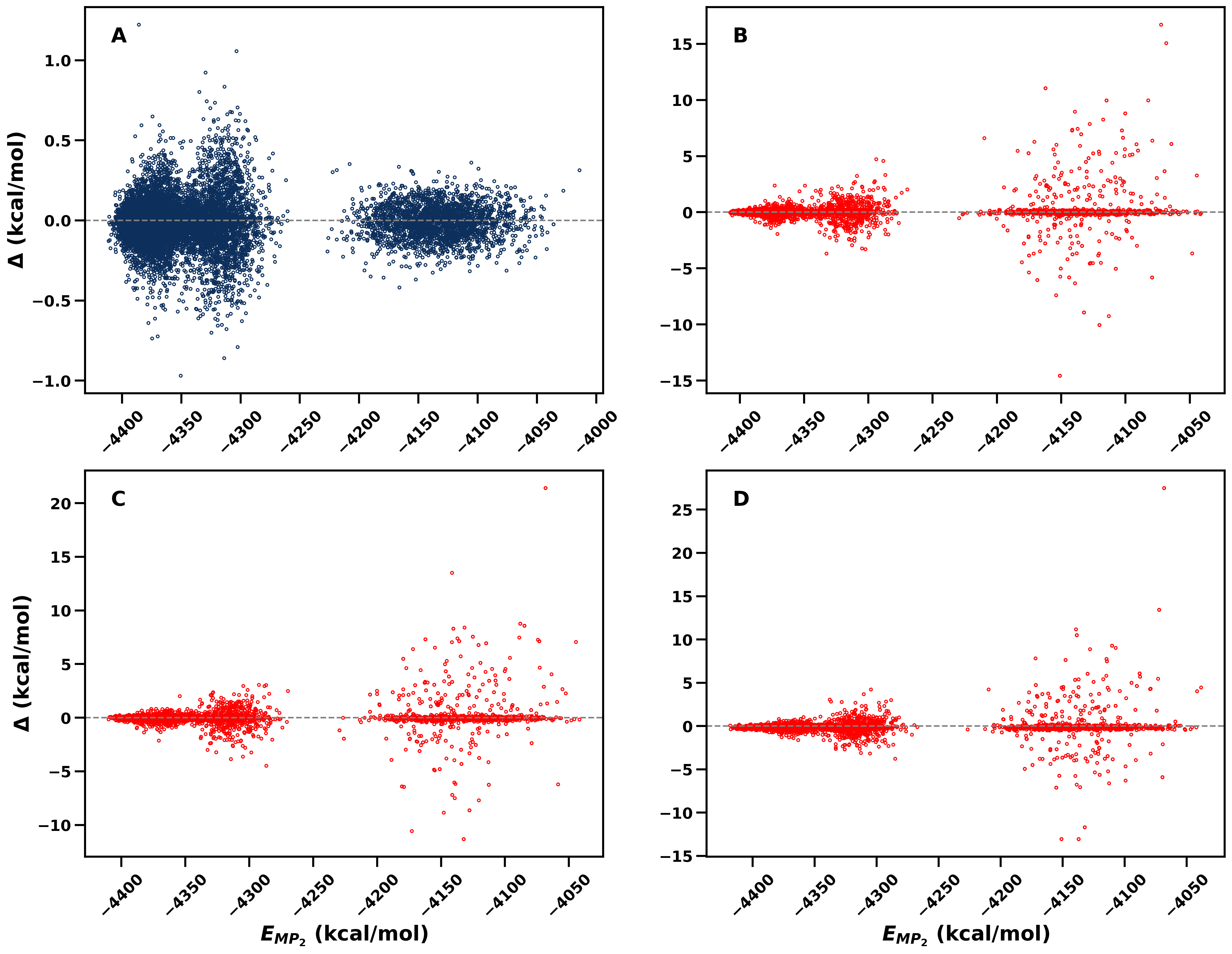}
\caption{Correlation of train (A) and test (B) set errors of SETG5, test set 
 errors of SETS1(C) and SETS2 (D) reference energies and predicted NN energies.}
\label{sifig:aka_error}
\hfill
\end{figure}

\begin{figure}[htbp]
\includegraphics[width=0.75\textwidth]{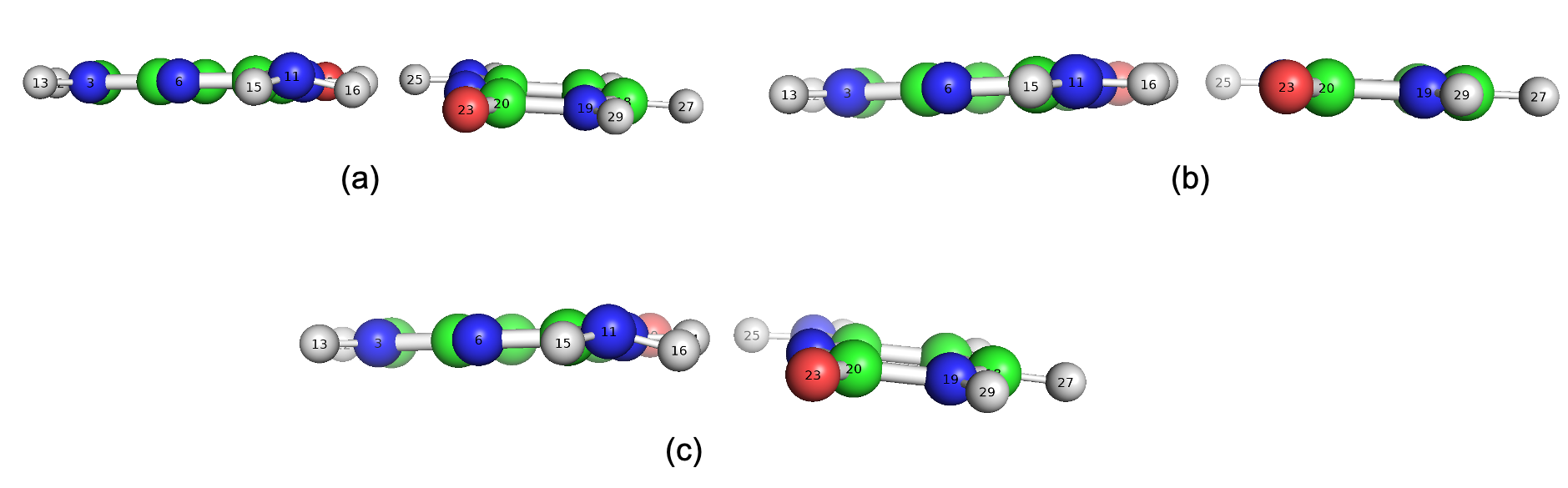}
\caption{Equilibrium Geometries of the GC pair obtained at different
  levels of theory: (a) MP2/cc-pVTZ; (b) B3LYP/def2-TZVP; (c) Geometry
  obtained from the PhysNet model after transfer learning to
  MP2/cc-pvTZ}
\label{sifig:eq.geom}
\hfill
\end{figure}

\begin{figure}[htbp]
\includegraphics[width=0.75\textwidth]{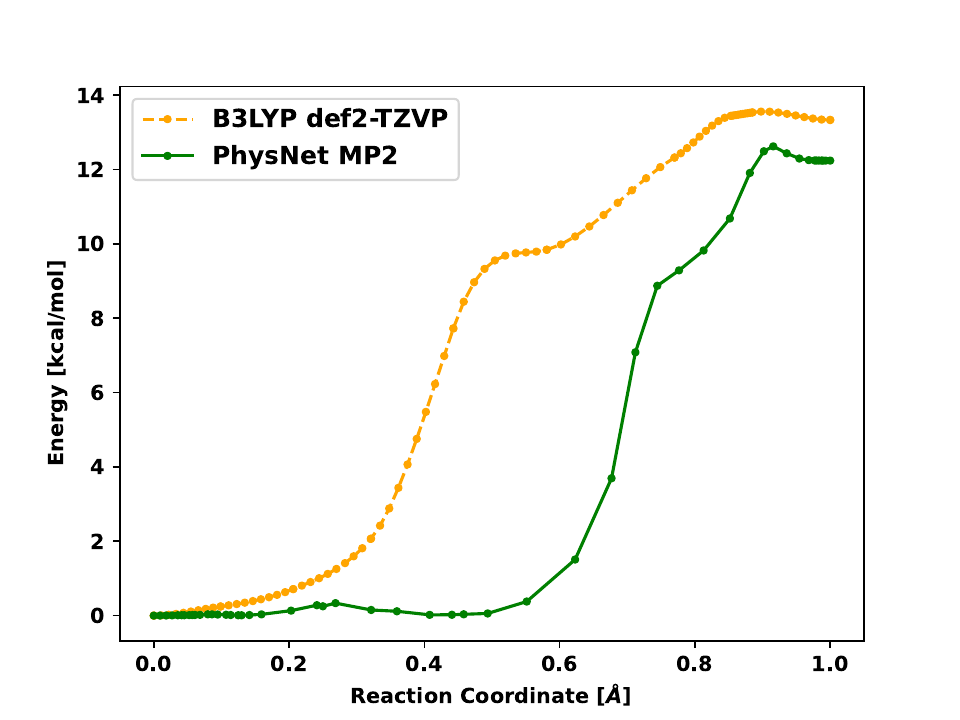}
\caption{Nudge Elastic Band reaction profiles for the AT system
  calculated at B3LYP/def2-TZVP level of theory and with the ML-PES
  transfer learned to the MP2 level of theory.}
\label{sifig:at_neb}
\hfill
\end{figure}

\begin{figure}[htbp]
\centering
\includegraphics[width=1.0\textwidth]{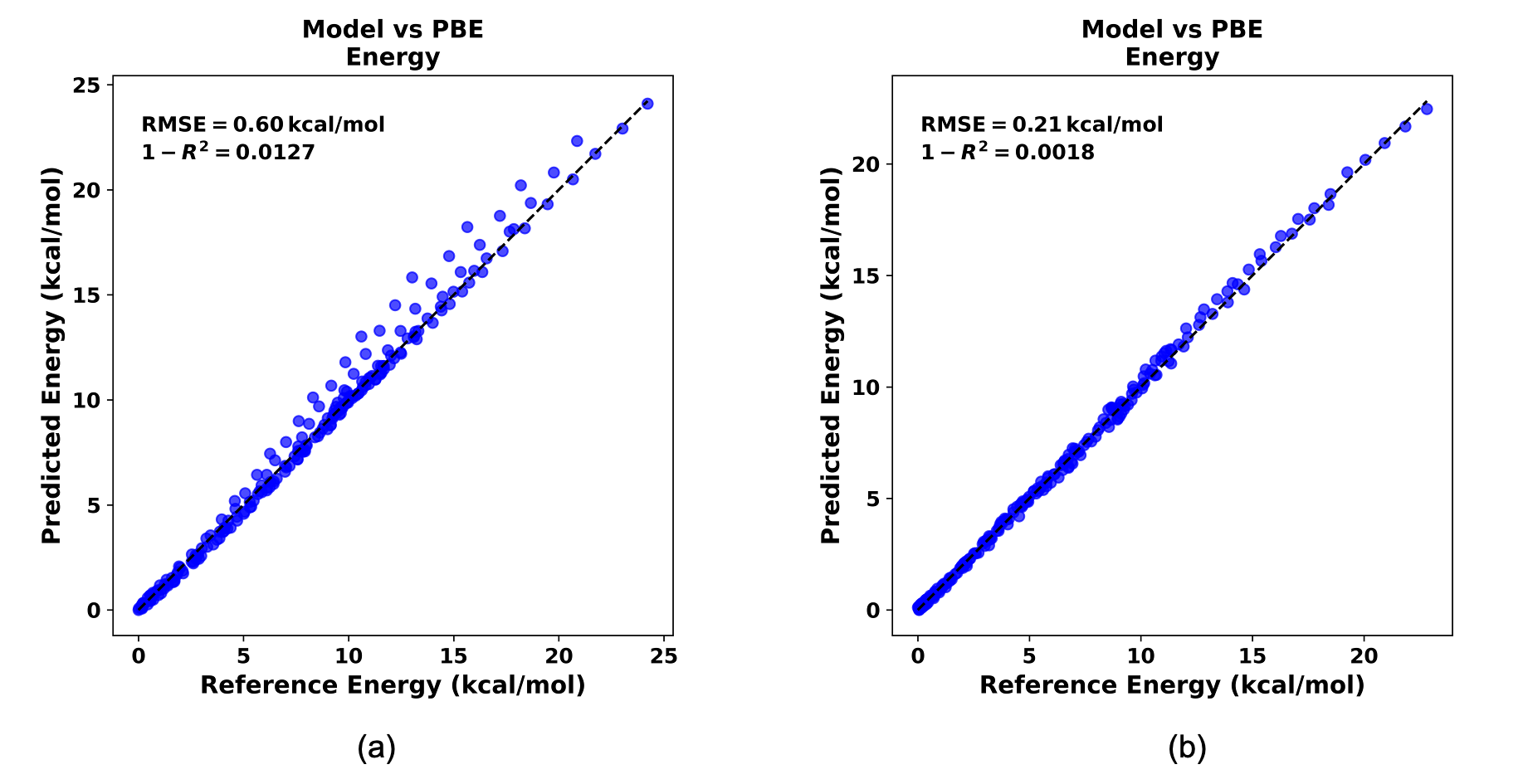}
\caption{Energy correlation plot for the energies in the range
  evaluated in Figure \ref{fig:fig3} of the main manuscript with the
  model constructed at the PBE/def2-SVP level of theory. Panel A shows
  the results for the GC pair, while Panel B shows the corresponding
  results for AT.}
\label{sifig:corr_plot_PBE}
\end{figure}

\begin{figure}[htbp]
\centering
\includegraphics[width=1.0\textwidth]{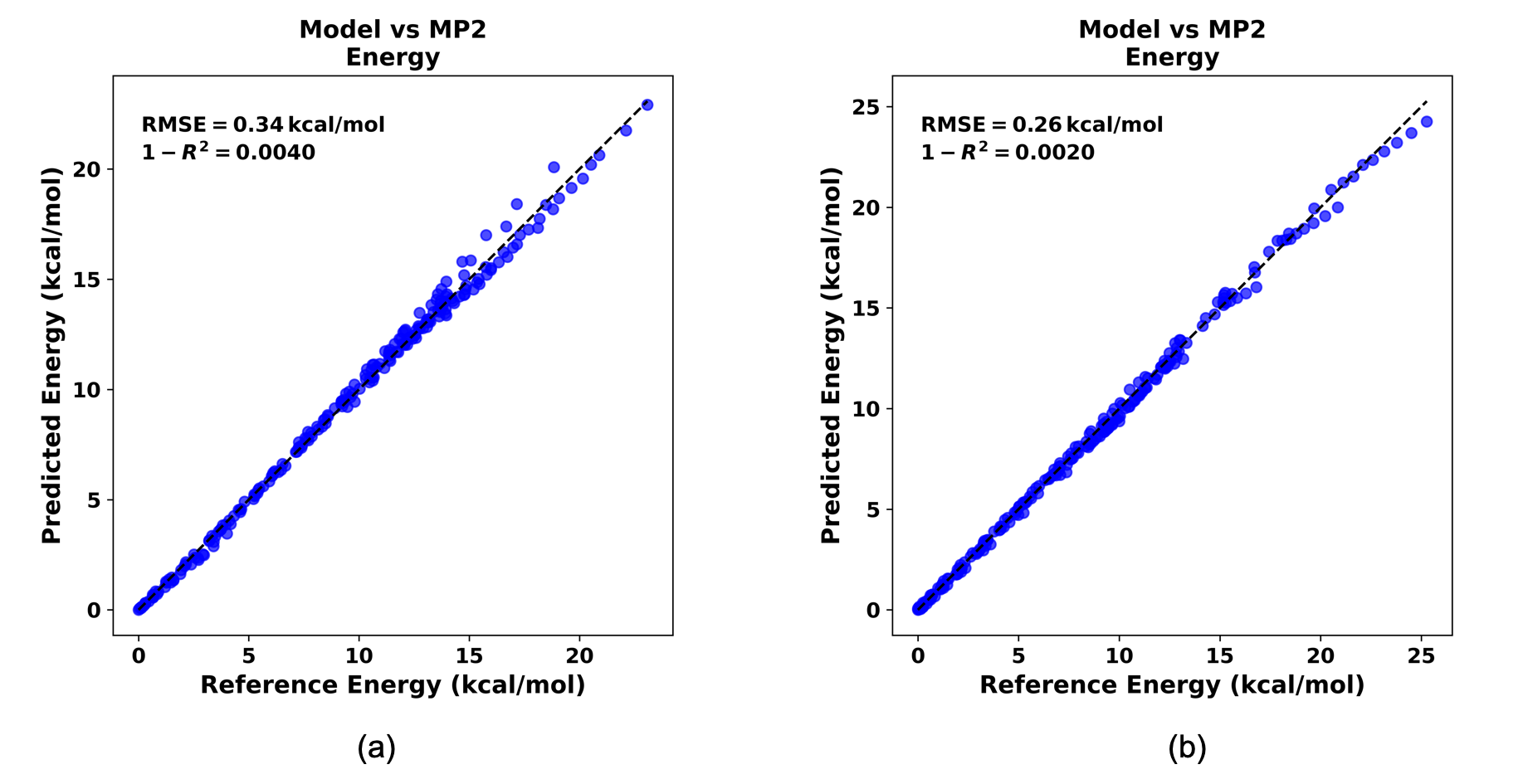}
\caption{Correlation between MP2/cc-pVTZ reference energies and the
 TL-PES retrained with 2300 points, for CG (panel A) and AT (panel B).}
\label{sifig:corr_plot_MP2}
\end{figure}

\begin{figure}[h!]
\centering
\includegraphics[width=1.0\textwidth]{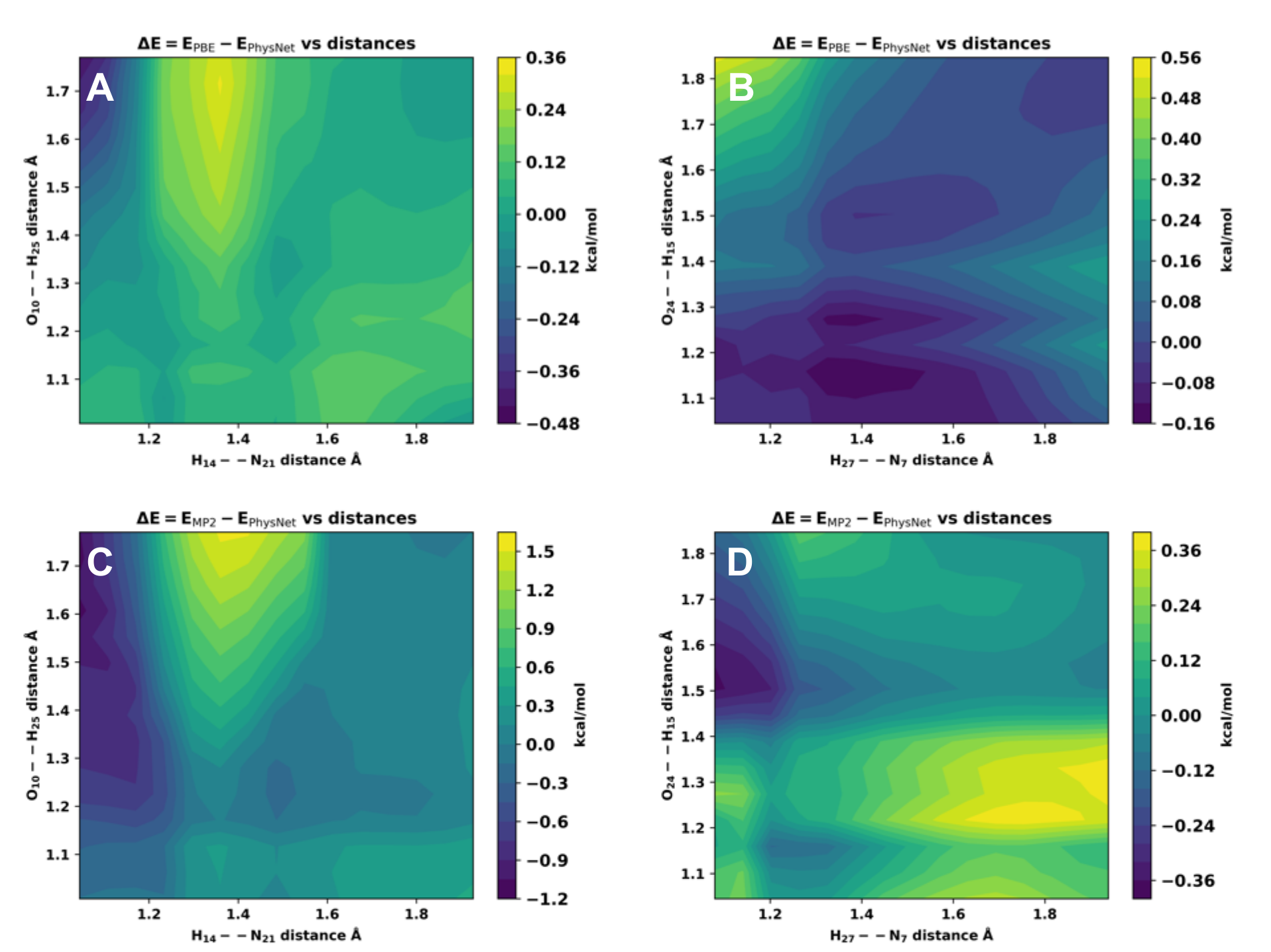}
\caption{Differences of energy values between reference PBE
  calculations and predicted Physnet values (Panels A/B). Panels C/D
  show the same for MP2 reference and Physnet predictions. Panels A/C
  are for GC basepair, while panels B/D are for AT basepair.}
\label{sifig:delta-maps}
\end{figure}

\begin{figure}[h!]
\centering
\includegraphics[width=1.0\textwidth]{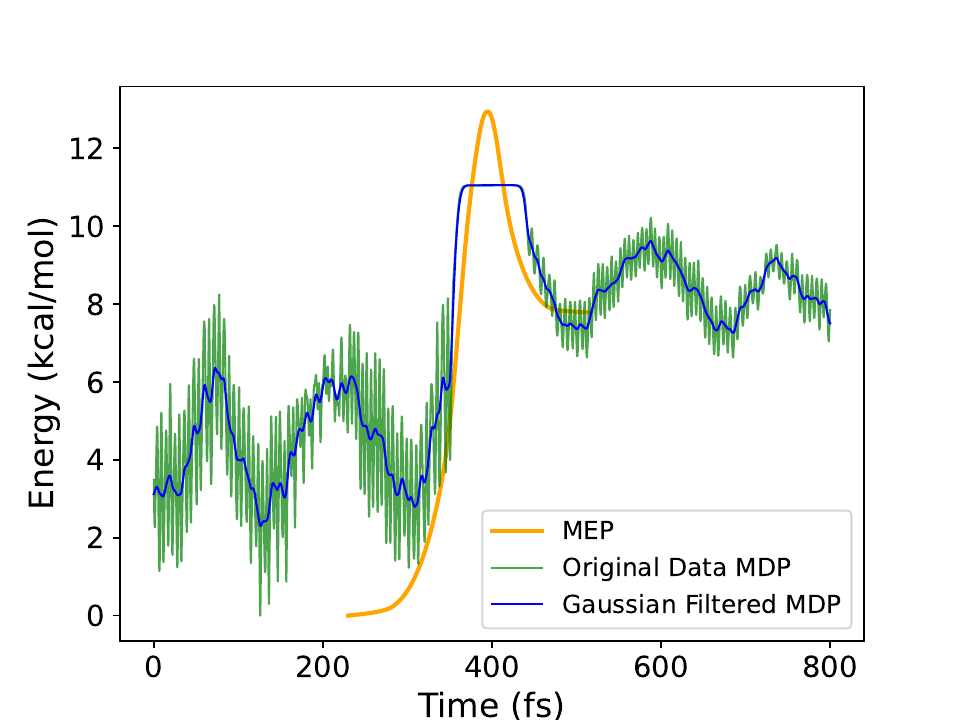}
\caption{Minimum Energy Profile and Minimum Dynamic Path for the GC dimer with the transfer ML-PES at the MP2 level.}
\label{sifig:mdp_mep}
\end{figure}

\begin{figure}[h!]
\centering
\includegraphics[width=1.0\textwidth]{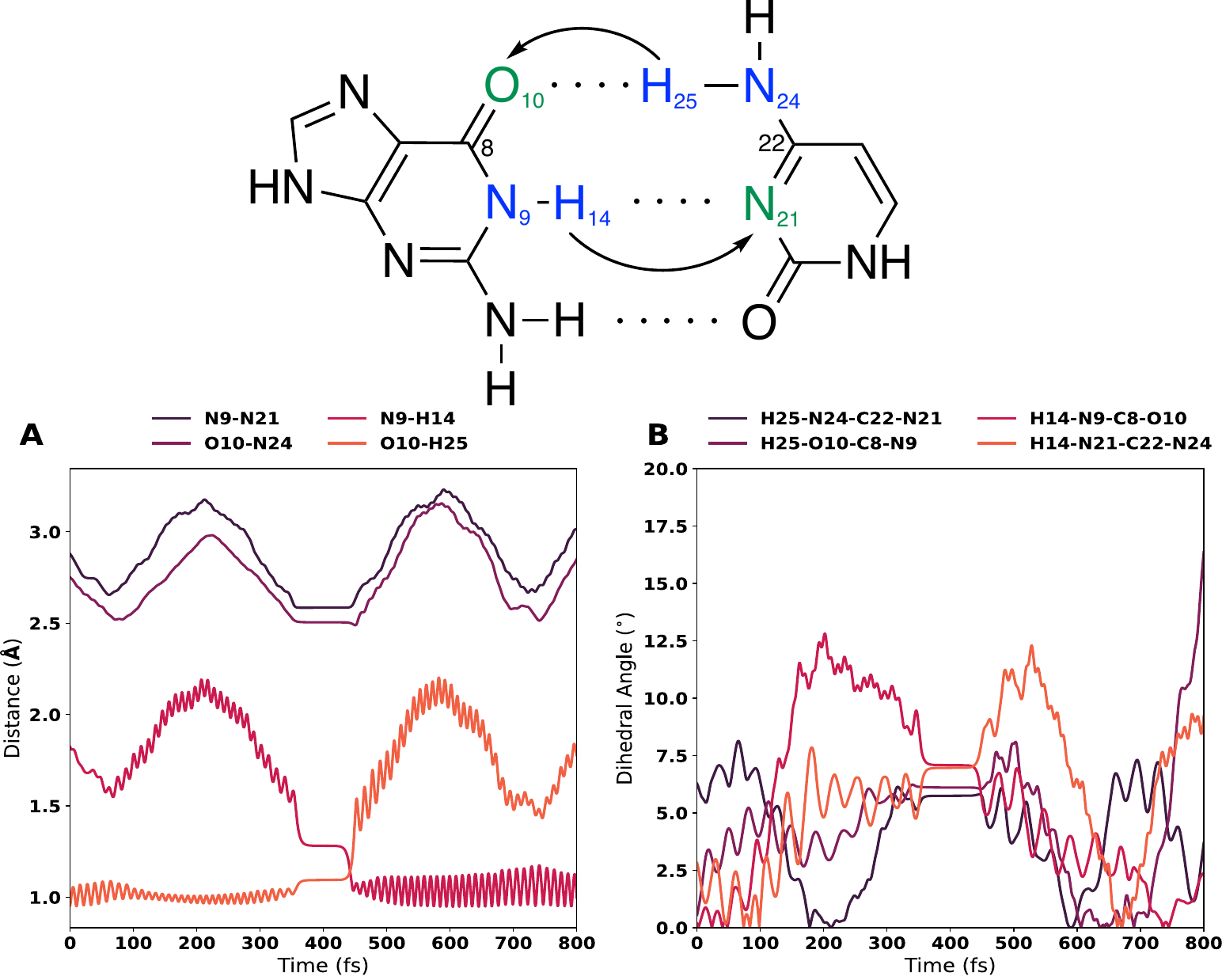}
\caption{Evolution of the internal degrees of freedom of the CG dimer over the minimum dynamic path for the transfer ML-PES at the MP2 level. Panel A shows the distances between the heavy atoms involved in the reaction (N9-N21, O10-N24), and the distance to the hydrogen atoms to the nearest heavy atom (N9-H14, O10-H25). Panel B shows the dihedral angles of the atoms involved in the reaction (
  H24-N23-C21-N20, H24-O9-C7-N8, H13-N8-C7-O9 and H13-N20-C21-N23). On top, the mapping of the atoms in the CG pair.}
\label{sifig:mdp_internal}
\end{figure}

\end{document}